\documentclass[article,notitlepage,amsthm,amsmath,amssymb,nofootinbib]{revtex4-2}
\usepackage[T1]{fontenc}
\usepackage{color}
\usepackage[dvipsnames]{xcolor}
\usepackage{graphicx}
\usepackage{amsmath}
\usepackage{amssymb}
\usepackage{amsthm}
\usepackage{longtable}
\usepackage[braket]{qcircuit}
\usepackage[colorlinks = true,linkcolor = blue,urlcolor = blue,citecolor = blue,anchorcolor = blue]{hyperref}
\hypersetup{hypertexnames=false,pdftitle={Depth optimization of CZ circuits and Clifford synthesis on the square lattice},pdfauthor={Dmitri Maslov}}

\newcommand{\Gate}[1]{\textnormal{\textsc{#1}}}
\newcommand{\idgate}{\Gate{I}}
\newcommand{\cnotgate}{\Gate{CNOT}}
\newcommand{\cygate}{\Gate{CY}}
\newcommand{\czgate}{\Gate{CZ}}
\newcommand{\hgate}{\Gate{H}}
\newcommand{\xgate}{\Gate{X}}
\newcommand{\ygate}{\Gate{Y}}
\newcommand{\zgate}{\Gate{Z}}
\newcommand{\pgate}{\Gate{P}}

\newcommand{\lcgate}{\Gate{LC}}

\newcommand{\calD}{{\cal D }}

\newcommand{\calL}{{\cal L }}

\newcommand{\calQ}{{\cal Q }}

\newcommand{\FF}{\mathbb{F}}
\newcommand{\ZZ}{\mathbb{Z}}
\DeclareMathOperator{\Span}{span}
\DeclareMathOperator{\rank}{rank}

\newtheorem{prop}{Proposition}
\newtheorem{lemma}{Lemma}
\newtheorem{cor}{Corollary}

\newcommand{\eq}[1]{\hyperref[eq:#1]{Eq.~(\ref*{eq:#1})}}
\renewcommand{\sec}[1]{\hyperref[sec:#1]{Section~\ref*{sec:#1}}}
\newcommand{\ssec}[1]{\hyperref[ssec:#1]{Subsection~\ref*{ssec:#1}}}
\newcommand{\fig}[1]{\hyperref[fig:#1]{Figure~\ref*{fig:#1}}}
\newcommand{\tab}[1]{\hyperref[tab:#1]{Table~\ref*{tab:#1}}}
\newcommand{\lem}[1]{\hyperref[lem:#1]{Lemma~\ref*{lem:#1}}}
\newcommand{\propos}[1]{\hyperref[propos:#1]{Proposition~\ref*{propos:#1}}}
\newcommand{\app}[1]{\hyperref[app:#1]{Appendix~\ref*{app:#1}}}

\newcommand{\ResultCount}{481}
\newcommand{\CountingCount}{479}
\newcommand{\SmallCount}{9}
\newcommand{\ACount}{29}
\newcommand{\PCount}{36}
\newcommand{\GmgCount}{152}
\newcommand{\LCount}{255}
\newcommand{\LargestN}{1{,}024}

\newcommand{\ContiguousN}{256}

\makeatletter
\let\auto@bib@innerbib\@empty
\let\write@bibliographystyle\relax
\makeatother

\begin{document}

\title{\texorpdfstring{Shallow Clifford circuits on square lattices}{Shallow Clifford circuits on square lattices}}
\author{Dmitri Maslov}
\affiliation{IonQ, 4505 Campus Dr, College Park, MD 20740, USA}
\date{November 6, 2026}

\begin{abstract}
We study the depth of $\czgate$ circuits, Clifford state preparation, and Clifford unitary synthesis over the two-dimensional square lattice.  A programmable circuit on a $2{\times}k$ ladder ($n{=}2k$) implements an arbitrary $\czgate$ computation in depth at most $\frac{3n}{2}{+}3$ up to reversal of both rails.  This improves the best-known LNN (Linear Nearest-Neighbor)-compliant depth-$(2n{+}2)$ construction by about 25\%.  For state preparation, we consider a Phase gate programmable -$\hgate$-\{$\cnotgate$+$\pgate$\}-$\lcgate$- circuit and derive matching obstructions to attaining the $\czgate$-span counting lower bound of $2\lceil\frac{n}{2}\rceil{-}1$ for depth by the entangling gates.  For cycle connectivity $C_n$, we prove that the minimum depth is exactly $2\lceil\frac{n}{2}\rceil{-}1$ when $n$ is not a multiple of four and $2\lceil\frac{n}{2}\rceil$ when $n\,{=}\,4k$.  We construct a hole-free minimum-perimeter square-lattice patch with a Hamiltonian cycle for every even size, giving optimal-depth programmable circuits to implement Clifford states for all $n\,{=}\,4k{+}2$.  Exhaustive and heuristic searches address the remaining cases for $n$ up to 256, and selected values up to 1,024 qubits.  The improvement against comparable state-of-the-art, an LNN-compliant depth-$(2n{+}2)$ solution, is about 50\%.   Finally, combining state synthesis circuits developed in this work with a known four-diagonal-stage decomposition gives rise to a depth $4n{-}4$ for $n\,{=}\,4k{+}2$ and depth $4n$ for $n\,{=}\,4k$ implementation of arbitrary Clifford unitaries.  This improves the most relevant state-of-the-art, an LNN-compliant depth $7n{-}4$ construction, by about 43\%.
\end{abstract}

\maketitle

\section{Introduction}\label{sec:Intro}

Clifford circuits play an important role in quantum computing.  They underlie quantum error correction \cite{gottesman1998theory} and state distillation \cite{bravyi2005universal,bravyi2012magic}, appear in randomized benchmarking \cite{knill2008randomized,magesan2011scalable} and quantum tomography \cite{kueng2016lowrank,huang2020predicting}, and provide a useful setting for the study of entanglement \cite{fattal2004entanglement,hein2004multiparty}.  Their efficient classical simulation \cite{aaronson2004improved} does not remove the need for efficient quantum circuit implementations: the quality and runtime of the applications depend on the circuits that are executed.  A substantial body of work concerns the structure, synthesis, and optimization of Clifford circuits \cite{aaronson2004improved,maslov2018shorter,bravyi2021hadamard, nemirovsky2026global, goubault2023shallower}.

Here, we focus on the depth by entangling gates.  This cost metric reflects the time to solution when single-qubit gates are inexpensive compared to entangling gates, and it is relevant when decoherence limits the quality of a computation.  A fault-tolerant quantum computation can also favor low depth by entangling gates for similar reasons.  The architecture is another important consideration.  Unrestricted connectivity allows circuit constructions that need not remain efficient after routing to a nearest-neighbor architecture.  Over unrestricted connectivity, an arbitrary $\cnotgate$ circuit admits depth $O(n/\log n)$ without ancillary qubits, with an asymptotically optimal space--depth trade-off when ancillary qubits are available \cite{jiang2020optimal}.  These bounds are based on the synthesis of a linear reversible transformation.  Conversely, a construction designed for a restricted architecture can use intermediate linear functions to implement several desired interactions without explicitly moving each interacting pair next to one another.

The two-dimensional square lattice (2DSL) is motivated by several hardware platforms.  Google's Sycamore and Willow superconducting processors use square grids of transmon qubits with nearest-neighbor couplings \cite{arute2019quantum,google2025belowthreshold}.  IonQ's proposed walking cat architecture for trapped ions uses a square-grid moving-qubit model in which two-qubit gates act between nearest-neighbor sites \cite{tripier2026walking}.  Neutral-atom processors have implemented gate-model algorithms with Rydberg-mediated $\czgate$ gates in two-dimensional square arrays \cite{graham2022multiqubit}.  These platforms differ in their approach to computational qubit implementation, but each utilizes a 2DSL architecture, explaining and motivating our focus.  Specifically, we study the local circuit synthesis problem on a fixed occupied square-lattice layout, with gates restricted to nearest neighbors.

The usefulness of intermediate linear functions is particularly apparent for Hadamard-free Clifford circuits.  Such circuits admit a phase polynomial description and can be written as a three-stage computation -$\pgate$-$\czgate$-$\cnotgate$- \cite{maslov2018shorter}.  The same work gives a linear nearest-neighbor (LNN) reversal circuit into which Phase gates can be inserted to implement an arbitrary -$\czgate$- stage in depth $2n{+}2$, together with qubit reversal.  Subsequent work optimized the depth of $\czgate$, $\cnotgate$, and Clifford circuits over unrestricted connectivity \cite{maslov2022depth}, and showed how $\czgate$ circuitry can be incorporated into LNN linear reversible circuits without increasing their worst-case depth \cite{maslov2023cnot}.  

Recent synthesis methods include phase polynomial algorithms respecting connectivity constraints \cite{vandaele2022phase}, graph-state frameworks for Clifford isometries \cite{goubault2025graph}, optimal, A*, and greedy algorithms for $\cnotgate$ and Clifford operators \cite{webster2026heuristic}, and graph-decimation methods for stabilizer states \cite{doherty2026fast}.  These approaches optimize specified transformations or states.  Here, the principal search objective is a fixed entangling circuit that can be programmed to supply every $\czgate$, Clifford state, or Clifford unitary computation.

Yang and Rall \cite{yang2024harnessing} study square-lattice graph-state and Clifford synthesis assisted by ancillary entanglement buses, counting depth in layers of GHZ-state injections.  We count nearest-neighbor two-qubit layers.  Our even-size constructions use no ancillary qubits or measurements; the odd-size unitary extension uses one additional qubit, returned unchanged.

Our first result extends the reversal-walk \cite{kutin2007computation} approach to a $2{\times}k$ ladder.  Two complementary walks make all within-rail \cite{maslov2018shorter} and cross-rail $\czgate$ terms programmable in depth at most $3k{+}3$.  The within-rail terms follow from the one-dimensional walk \cite{maslov2018shorter}.  The new part is a proof that the rung interactions span all $k^2$ cross-rail terms for every $k$.  This gives a constructive linear-depth upper bound with a simple endpoint qubit order reversal.

Our second focus is the preparation of Clifford states, also called pure stabilizer states.  Every such state is local-Clifford equivalent to a graph state \cite{vanden2004graphical}.  This allows us to study a universal programmable circuit whose entangling gates are selected once and whose single-qubit Phase gates specify the desired $\czgate$ computation.  We seek circuits of the form -$\hgate$-\{$\cnotgate$+$\pgate$\}-$\lcgate$-.  The initial -$\hgate$- stage contains a Hadamard gate on every qubit, while the final -$\lcgate$- stage contains arbitrary single-qubit Clifford gates.  We rely on the notion that the \{$\cnotgate{+}\pgate$\} part may have any fixed linear invertible endpoint (suffix/prefix); its inverse is unnecessary for state preparation.  The goal here is to construct a depth-optimal 2DSL-compliant programmable \{$\cnotgate{+}\pgate$\} circuit.

The distinction between a fixed universal programmable circuit and synthesis of one desired state is important.  We require the middle stage to span the entire $\czgate$ space, with local phases available separately.  The depth lower bound, the algorithmically constructed circuits, and the searched results refer to this programming requirement.  The final local Clifford gates give universality for Clifford states; this does not imply that a full $\czgate$ span is necessary for each state-universal template exploiting those local gates, or that each individual state requires the reported depth.

The connectivity itself also constrains an optimal-depth circuit: every layer must be a maximum matching and every gate must contribute an independent CZ dimension via Phase gate insertion.  We derive matching obstructions, exclude all trees on at least four vertices, and show why small diameter alone does not guarantee success.  In contrast, a simple cycle reaches the bound unless its size is divisible by four, when precisely one additional layer is required.  We prove that every even size admits a compact, hole-free minimum-perimeter square-lattice patch containing a Hamiltonian cycle.  This gives an explicit optimal-depth state template for every $n\,{=}\,4k{+}2$, while searches retain their role for odd sizes and for improving the cycle depth at $n\,{=}\,4k$.  The minimal-perimeter hole-free requirement is selected to make sure the patch implementing the Clifford state/unitary is compact in the 2DSL.

Our final result concerns arbitrary Clifford unitaries (in the strict sense: no permutation of output qubits allowed).  A programmed full-span circuit and its binary transpose can supply two alternating diagonal stages while cancelling their linear endpoints (suffix/prefix).  The even-cycle construction has the transpose symmetry needed for this pairing.  Combining it with the four-stage decomposition of Nemirovsky et al.\ \cite{nemirovsky2026global} gives depths $4n{-}4$ and $4n$ in the two even congruence classes, $n\,{=}\,4k{+}2$ and $n\,{=}\,4k$, correspondingly.  This result holds on any even-length Hamiltonian cycle in 2DSL, including over the minimum-perimeter patches constructed here.

The paper is organized as follows.  \sec{ladder} describes the $2{\times}k$ construction.  \sec{states} develops the state-synthesis reduction, rank criterion, lower bounds, and connectivity results, including cycles and their square-lattice embeddings.  \sec{search} describes the search algorithms and their exact pruning rules.  \sec{results} reports certified results from a fixed computational snapshot.  \sec{unitaries} gives the Clifford unitary construction.  The ladder span proof, explicit small circuits, and direct proof of the cycle construction appear in \app{ladder}, \app{small}, and \app{cycles}, correspondingly.

\subsection{Notation and circuit model}\label{ssec:notation}

We use the gate library consisting of the Hadamard, Phase, $\cnotgate$, and $\czgate$ gates,
\begin{equation*}
\hgate\,{:=}\,\frac{1}{\sqrt{2}}\begin{pmatrix}1&1\\1&{-}1\end{pmatrix},\qquad
\pgate\,{:=}\,\begin{pmatrix}1&0\\0&i\end{pmatrix},\qquad
\cnotgate\ket{a,b}\,{:=}\,\ket{a,a\oplus b},\qquad
\czgate\ket{a,b}\,{:=}\,({-}1)^{ab}\ket{a,b}.
\end{equation*}
Here, $a,b\,{\in}\,\{0,1\}$ and $\oplus$ denotes addition over $\FF_2$.  In particular, $\pgate^2\,{=}\,\zgate$ and $\pgate^3\,{=}\,\pgate^\dagger$.  We use $\pgate$ for the Phase gate throughout.  $\lcgate$ denotes a tensor product of single-qubit Clifford gates generated by $\hgate$ and $\pgate$.  Global phases are ignored.

The physical qubits occupy distinct sites of the square lattice.  A two-qubit gate may act only between sites at Manhattan distance one, and no qubit participates in more than one two-qubit gate in the same layer.  The connectivity results also consider general graphs, with gates restricted to their edges.  There are no ancillary qubits or measurements, except for the explicitly stated one-additional-qubit extension for odd-size unitary synthesis.  Unless explicitly stated otherwise, depth counts only two-qubit layers.  Single-qubit Phase gates and the initial and final local Clifford stages do not contribute to this metric.  Circuits are read from left to right; operator products act on state vectors from right to left.

Clifford operations and stabilizer states admit a description in terms of binary linear maps and quadratic forms \cite{dehaene2003clifford}.  For input variables $x_1,x_2,...,x_n$ and a symmetric zero-diagonal binary matrix $\Gamma$, define
\begin{equation}
D_\Gamma\,{:=}\,\prod_{1\,{\leq}\,i\,{<}\,j\,{\leq}\,n}\czgate(i,j)^{\Gamma_{ij}},\qquad
D_\Gamma\ket{x_1,x_2,...,x_n}
\,{=}\,({-}1)^{\sum_{i\,{<}\,j}\Gamma_{ij}x_ix_j}\ket{x_1,x_2,...,x_n}.\nonumber
\end{equation}
The above $n$-qubit $\czgate$ circuits form a linear space over $\FF_2$ of dimension $D(n)\,{:=}\,\binom{n}{2}$: identify $D_\Gamma$ with its coefficient vector $(\Gamma_{ij})_{i\,{<}\,j}\,{\in}\,\FF_2^{D(n)}$.  All $\czgate$ gates commute and square to the identity, so circuit composition corresponds to modulo-2 vector addition, $D_\Gamma D_\Lambda\,{=}\,D_{\Gamma{\oplus}\Lambda}$.  The identity circuit is the zero vector, and the gates $\czgate(i,j)$, $1\,{\leq}\,i\,{<}\,j\,{\leq}\,n$, are the standard basis vectors \cite{maslov2023cnot}.  We refer to the coordinates of this space as $\czgate$ dimensions.  

We use the phase polynomial terminology of \cite{maslov2018shorter}.  For two basic input variables $a,b$,
\begin{equation}\label{eq:phasepair}
\czgate(a,b)\,{=}\,\pgate(a\,{\oplus}\,b)\,\pgate^\dagger(a)\,\pgate^\dagger(b) \,{\sim}\, \pgate(a\,{\oplus}\,b).
\end{equation}
The last part is due to the availability of primary inputs in our constructions.  Thus, a Phase gate on the EXOR supplies the desired $\czgate$.  More generally, a Phase gate on the EXOR of a collection of basic variables supplies all pairwise $\czgate$ gates.

For EXORs $u,v$ of input variables, we use $\czgate(u,v)$ for the phase operation on the right-hand side of \eq{phasepair}.  In particular, for disjoint lists $a_1,a_2,...,a_r$ and $b_1,b_2,...,b_s$,
\begin{equation}\label{eq:rectangle}
\czgate\left(\bigoplus_{i=1}^{r}a_i,\ \bigoplus_{j=1}^{s}b_j\right)
\,{=}\,\prod_{i=1}^{r}\prod_{j=1}^{s}\czgate(a_i,b_j).
\end{equation}
This is a rectangle of $\czgate$ gates between the two lists, as in \cite[Lemma 1]{maslov2022depth}: every row variable interacts with every column variable.  In the coefficient space above, composition is addition and repeated gates cancel.  If the EXORs share variables, identical arguments contribute only an input Phase correction.  Thus, in this coefficient space,
\begin{equation*}
\czgate(u\,{\oplus}\,u',v)\,{=}\,\czgate(u,v)\,\czgate(u',v),\qquad
\czgate(u,v)\,{=}\,\czgate(v,u),\qquad \czgate(u,u)\,{=}\,0.
\end{equation*}
An invertible change of the input variables preserves the dimension of this $\czgate$ coefficient space: the substitution and its inverse both send a linear Pauli-$\zgate$ phase to another linear Pauli-$\zgate$ phase.

\section{\texorpdfstring{$\czgate$ circuits on a $2{\times}k$ lattice}{CZ circuits on a 2 by k lattice}}\label{sec:ladder}

Consider a ``ladder'' ($2{\times}k$ lattice) with input variables $a_1,a_2,...,a_k$ on rail A and $b_1,b_2,...,b_k$ on rail B.  Horizontal edges join consecutive positions on the same rail; the $j^{\rm th}$ rung joins the two positions numbered $j$.  We construct a fixed horizontal $\cnotgate$ circuit and allow Phase gates on the rails and independently selected $\czgate$ gates on the rungs.  Thus, this construction uses both $\cnotgate$ and $\czgate$ as entangling gates.  

\begin{figure}[t]
\centering
\includegraphics[width=0.88\textwidth]{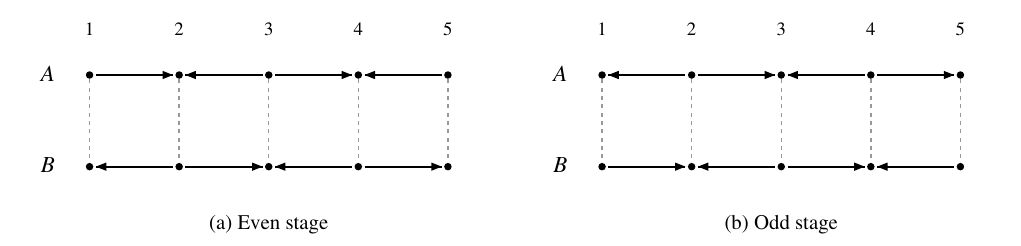}
\caption{Complementary horizontal stages on a $2{\times}5$ ladder.  Arrows point from $\cnotgate$ controls to targets.  Each displayed horizontal stage is scheduled in two matching layers.  Dashed vertical edges denote the programmable rung $\czgate$ gates applied after the horizontal stage, in one additional layer.  Stage parity refers to $t\,{=}\,0,1,...,k$.}
\label{fig:ladder}
\end{figure}

At stage $t$, $t\,{=}\,0,1,...,k$, apply the following ``horizontal'' gates (in depth 2):
\begin{enumerate}
\item On rail A, every position $j\,{\equiv}\,t\pmod{2}$ receives a $\cnotgate$ from each of its existing neighbors.
\item On rail B, every position $j\,{\not\equiv}\,t\pmod{2}$ receives a $\cnotgate$ from each of its existing neighbors.
\end{enumerate}
All controls on a rail have one checkerboard color and all targets the other.  The two gate layers commute.  Partitioning horizontal edges by the parity of their left endpoint schedules a stage in at most two matching layers, with the two rails executed in parallel.  After each stage, allow one independently programmable $\czgate$ on every rung.  All rungs form a matching, so this costs one further two-qubit layer.  Phase gates may be inserted at the stage boundaries, including the input boundary.  \fig{ladder} illustrates the two alternating horizontal stages.

Let $A_t$ and $B_t$ be the binary matrices describing the rail parities after $t$ horizontal stages, with $A_0\,{=}\,B_0\,{=}\,I$.  Denote their $j^{\rm th}$ rows by $\alpha_j^{(t)}$ and $\beta_j^{(t)}$, correspondingly.  A selected rung gate at boundary $t$ supplies the rectangle
\begin{equation}
\czgate(\alpha_j^{(t)}a,\beta_j^{(t)}b),\qquad
M_{t,j}\,{:=}\,(\alpha_j^{(t)})^T\beta_j^{(t)}.\nonumber
\end{equation}
The matrix $M_{t,j}\,{\in}\,\FF_2^{k\times k}$ records its cross-rail $\czgate$ coefficients in the input variables.  Addition of these matrices corresponds to composition of rung phase contributions.

\begin{lemma}\label{lem:AB}
For every $k\,{\geq}\,1$, the complementary ladder walks satisfy
\begin{equation}
\Span_{\FF_2}\{M_{t,j}:1\,{\leq}\,t\,{\leq}\,k{+}1,\ 1\,{\leq}\,j\,{\leq}\,k\}
\,{=}\,\FF_2^{k\times k}.\nonumber
\end{equation}
Their endpoints are $A_{k{+}1}\,{=}\,B_{k{+}1}\,{=}\,J_k$, where $J_k$ reverses the $k$ positions.
\end{lemma}

The direct proof in \app{ladder} changes the input variables on each rail so that the selected rung slots become squares ($\czgate$ rectangles with equal sides, see \fig{ladderrectangles} and \fig{ladderrectanglesfive} for illustration) on matching cyclic intervals and on intervals shifted by one position.  Four nested squares cancel to leave two opposite corners.  Canceling the corner at a smaller index difference then isolates one $\czgate$ gate at a time, until every cross-rail pair is obtained.  The $2{\times}4$ example in \fig{ladderrectangles} illustrates this cancellation.

For the corollary below, observe also that every horizontal stage is its own inverse, and its complementary stage is its transpose.  Thus $B_t\,{=}\,A_t^{-T}$, and the rung coefficient matrices satisfy
\begin{equation}\label{eq:projection}
M_{t,j}\,{=}\,A_t^T E_{jj}A_t^{-T}.
\end{equation}

\begin{cor}\label{cor:rungs}
The $k$ rung slots at the first sampled boundary, together with any $k{-}1$ rung slots at each of the remaining $k$ boundaries, form a basis of the cross-rail $\czgate$ space.  More generally, any $s$ sampled boundaries, $1\,{\leq}\,s\,{\leq}\,k{+}1$, have rank $1{+}s(k{-}1)$.
\end{cor}

\begin{proof}
By \eq{projection}, $\bigoplus_{j\,{=}\,1}^k M_{t,j}\,{=}\,I$ at every boundary.  Equality of the $k{+}1$ layer sums gives $k$ independent relations among the $k(k{+}1)$ indexed slots.  \lem{AB} shows that their nullity is exactly $k$, so these are all the relations.  Restricting to $s$ boundaries leaves $s{-}1$ independent relations.  Omitting one slot at every boundary but the first removes all nonzero relations among the retained slots.
\end{proof}

The horizontal walks also supply the within-rail $\czgate$ terms through Phase insertion, as in the LNN reversal construction \cite{maslov2018shorter}, resulting in the following lemma.

\begin{lemma}\label{lem:ladderdepth}
For $n\,{=}\,2k$, an arbitrary $\czgate$ circuit can be implemented on a $2{\times}k$ lattice, together with simultaneous reversal of both rails, in two-qubit depth at most
\begin{equation}
3(k{+}1)\,{=}\,3k{+}3\,{=}\,\frac{3n}{2}{+}3.\nonumber
\end{equation}
\end{lemma}

\begin{proof}
The horizontal walks expose every contiguous interval EXOR on each rail at the allowed Phase boundaries, as proved in \app{ladder}.  On rail A, change to the prefix variables $y_j\,{=}\,a_1\,{\oplus}\,a_2\,{\oplus}\,...\,{\oplus}\,a_j$, with $y_0\,{=}\,0$.  Every interval is $a_i\,{\oplus}\,a_{i+1}\,{\oplus}\,...\,{\oplus}\,a_j\,{=}\,y_{i-1}\,{\oplus}\,y_j$.  Hence the available Phase slots include every $y_j$ and every $y_i\,{\oplus}\,y_j$.  By \eq{phasepair}, these supply all within-rail $\czgate(y_i,y_j)$ gates and their individual Phase corrections, as in \cite{maslov2018shorter}.  The prefix change is invertible, so every within-rail $\czgate$ circuit in the original variables is programmable.  The same argument applies to rail B.

\lem{AB} supplies every cross-rail $\czgate(a_i,b_j)$.  The two within-rail spaces and the cross-rail space involve disjoint pairs of input variables and have total dimension
\begin{equation*}
2\binom{k}{2}{+}k^2\,{=}\,\binom{2k}{2}.
\end{equation*}
There are $k{+}1$ stages, each with two horizontal matching layers and one vertical matching layer, for a total of $3k{+}3$ circuit layers.  The qubits are returned in reversed order on both rails.  %
\end{proof}

\section{\texorpdfstring{Clifford state synthesis and $\czgate$ programming}{Clifford state synthesis and CZ programming}}\label{sec:states}

\subsection{\texorpdfstring{Reduction to a $\cnotgate{+}\pgate$ circuit}{Reduction to a CNOT+P circuit}}\label{ssec:reduction}

Every pure Clifford state can be written as
\begin{equation}\label{eq:graphstate}
\ket{\psi}\,{=}\,L\ket{G}\,{=}\,L D_\Gamma\hgate^{\otimes n}\ket{0}^{\otimes n},
\end{equation}
where $L$ is local Clifford and $\Gamma$ is the adjacency matrix of a simple graph \cite[Theorem 1]{vanden2004graphical}.  Thus, implementing an arbitrary $\czgate$ stage, followed by local Clifford gates, suffices for Clifford state synthesis.  The graph is a logical interaction graph and need not be a subgraph of the physical square lattice, which of course poses a problem in direct implementation should only the square lattice connectivity be available.

Consider a fixed $\cnotgate$ circuit $C$.  Let $A\,{\in}\,\FF_2^{n\times n}$ describe its endpoint, and write $U_A\ket{x}\,{=}\,\ket{Ax}$.  Inserting Phase gates at prescribed slots gives an operator
\begin{equation}
U_{C,\theta}\ket{x}\,{=}\,i^{p_\theta(x)}\ket{Ax},\qquad
p_\theta(x)\,{=}\,\sum_{s\,{=}\,1}^{m}\theta_s\ell_s(x)\pmod{4},\nonumber
\end{equation}
where $\ell_s$ is the Boolean linear function present at slot $s$ and $\theta_s\,{\in}\,\ZZ_4$.  This is the phase polynomial description of $\cnotgate{+}\pgate$ circuits \cite{maslov2018shorter}.  We allow input Phase gates as well, so phases linear in the original variables are independently programmable.

Suppose the available slots can express every diagonal Clifford operator $D$.  Given the desired $D_\Gamma$ from \eq{graphstate}, select the program for
\begin{equation}\label{eq:conjugateprogram}
D\,{=}\,U_A^{-1}D_\Gamma U_A.
\end{equation}
Conjugation by a linear reversible operation preserves the diagonal Clifford group.  Since $U_A\ket{+}^{\otimes n}\,{=}\,\ket{+}^{\otimes n}$ (indeed, $\ket{+}^{\otimes n}$ is an eigenvalue-1 eigenstate of any reversible circuit, including linear reversible), we obtain
\begin{equation}\label{eq:endpointabsorb}
L U_{C,\theta}\hgate^{\otimes n}\ket{0}^{\otimes n}
\,{=}\,L U_A D\ket{+}^{\otimes n}
\,{=}\,L D_\Gamma\ket{+}^{\otimes n}\,{=}\,\ket{\psi}.
\end{equation}
This establishes the desired -$\hgate$-\{$\cnotgate$+$\pgate$\}-$\lcgate$- form, and explains why any leftover linear reversible computation does not matter.  In particular, the $\cnotgate$ circuit need not finish at the identity, or even at a qubit permutation.

It is important to program \eq{conjugateprogram}, rather than use the coefficients of $\Gamma$ unchanged in input coordinates.  Substituting $Ax$ into a quadratic Boolean function can additionally create linear terms, because $x_i^2\,{=}\,x_i$.  The resulting diagonal Clifford is still programmable, but these linear terms must be included.  An equivalent implementation convention reverses the entire $\cnotgate$-and-Phase schedule.  Under this convention, a phase insertion after layer $t$ moves to the position after layer $d{-}t$, including cut zero, while each $\cnotgate$ keeps its original control and target.  

The same construction also gives destructive measurement in an arbitrary stabilizer basis, i.e., the joint eigenbasis of $n$ independent commuting Pauli operators.  Such a basis can be written as $\{L D_\Gamma\hgate^{\otimes n}\ket{z}:z\,{\in}\,\FF_2^n\}$, with $L$ and $\Gamma$ fixed for the basis.  For the program in \eq{conjugateprogram}, the binary Fourier identity \eq{fouriertranspose} gives
\begin{equation*}
L U_{C,\theta}\hgate^{\otimes n}\ket{A^{\mathsf T}z}
\,{=}\,L D_\Gamma\hgate^{\otimes n}\ket{z}.
\end{equation*}
Execute the compiled state-preparation circuit backward, inverting every gate, and measure all qubits in the computational basis.  The outcome $y$ corresponds to the stabilizer-basis label $z\,{=}\,A^{-T}y$.  This invertible classical relabeling gives the desired basis measurement with the same entangling depth as state preparation and without ancillary qubits.  Eigenvalues of a different choice of generators follow by a further known classical change of labels.  One area of application is implementing the final recovery-and-readout stage \cite{bravyi2021hadamard} of Clifford randomized benchmarking \cite{knill2008randomized, magesan2011scalable}---in this section, we show this can be done in theoretically minimal depth (in the sense of \lem{depthbound}) over a compact 2DSL patch for effectively all numbers of qubits one can expect to encounter (\tab{allresults}).

The graph representative in \eq{graphstate} is not unique.  Local complementation at a graph vertex complements the graph induced by its neighbors and corresponds to local Clifford operations \cite{vanden2004graphical}.  A target-dependent synthesis algorithm may exploit this freedom.  For unrestricted connectivity, optimizing over LC-equivalent graph representatives gives minimum $\czgate$ count and preparation depth for small graph states \cite{cabello2011optimal}.  Recent methods based on integer linear programming and simulated annealing minimize graph edge count within LC classes \cite{sharma2026minimising}.  Edge count and entangling depth over restricted architectures are, however, distinct cost metrics.

In fact, we applied the Gmg algorithm discussed later in this paper to the problem of specific state synthesis, and found some small yet consistent reductions against the $d_0(n)$ bound for a universal programmable circuit.  %

\subsection{A rank criterion}\label{ssec:rank}

Track the Boolean linear function on wire $j$ by a row vector $w_j\,{\in}\,\FF_2^n$.  Initially, $w_j\,{=}\,e_j$, and a $\cnotgate$ with control $c$ and target $t$ updates only
\begin{equation}\label{eq:wireupdate}
w_t\,{\leftarrow}\,w_t\,{\oplus}\,w_c.
\end{equation}
For $w\,{\in}\,\FF_2^n$, define its $\czgate$ vector \cite{maslov2023cnot}
\begin{equation}\label{eq:quadraticvector}
q(w)\,{:=}\,(w_iw_j)_{1\,{\leq}\,i\,{<}\,j\,{\leq}\,n}\,{\in}\,\FF_2^{D(n)}.\nonumber
\end{equation}
Pair coordinates are ordered lexicographically, beginning with $(1,2),(1,3),...,(1,n),(2,3),...,(n{-}1,n)$.  Let $\calQ(C)$ be the list of $\czgate$ vectors at all allowed Phase slots, with repetitions retained when useful for indexing.

The rank criterion follows from the integer expansion of a Boolean parity,
\begin{equation}\label{eq:parityexpand}
\ell_w(x)\,{:=}\,\bigoplus_{j:w_j\,{=}\,1}x_j
\,{\equiv}\,\sum_{j\,{=}\,1}^n w_jx_j
{+}2\sum_{1\,{\leq}\,i\,{<}\,j\,{\leq}\,n}w_iw_jx_ix_j\pmod{4}.
\end{equation}
Terms of degree at least three have coefficients divisible by four.  Consequently, an odd power of $\pgate$ at this slot contributes $q(w)$ to the $\czgate$ coefficient vector, together with a linear phase.  An even power contributes only a Pauli-$\zgate$, which can be pushed to an end of a Clifford circuit and thus is ignored. Note that $\pgate^\dagger\,{=}\,\pgate^3\,{=}\,\pgate.\zgate$, and thus $\pgate^\dagger$ is unnecessary as a separate gate.  We thus discuss only the $\pgate$ gates.

\begin{prop}\label{propos:rank}
With independently programmable input Phase gates, a fixed $\cnotgate$ circuit $C$ implements every diagonal Clifford phase polynomial, up to its fixed endpoint $U_A$, if and only if
\begin{equation}\label{eq:fullrank}
\rank_{\FF_2}\calQ(C)\,{=}\,D(n).
\end{equation}
In particular, \eq{fullrank} suffices for universal Clifford state preparation in the form \eq{endpointabsorb}.
\end{prop}

\begin{proof}
By \eq{parityexpand}, the $\czgate$ coefficient vector of any program lies in the span of $\calQ(C)$, proving necessity.  For sufficiency, choose a basis from $\calQ(C)$ and solve a binary linear system expressing the desired $\czgate$ coefficient vector as a sum of these basis vectors.  Apply $\pgate$ at the corresponding slots.  The resulting polynomial has the correct quadratic part, and its remaining discrepancy is linear over $\ZZ_4$.  Input powers of $\pgate$ correct this discrepancy independently on $x_1,x_2,...,x_n$.
\end{proof}

For example, $\pgate$ on the parity $x_1\,{\oplus}\,x_2$ supplies the exponent $x_1{+}x_2{+}2x_1x_2$.  Thus $\pgate(x_1\,{\oplus}\,x_2)\,{\sim}\,\czgate(x_1,x_2)$, as in \eq{phasepair}.  The same calculation explains why a binary rank computation must be accompanied by bookkeeping of linear phases when extracting an actual program.

For incoming control and target rows $u,v$, the gate increment can be expressed using the $\czgate$ vector already defined:
\begin{equation}\label{eq:increment}
q(u\,{\oplus}\,v)\,{\oplus}\,q(u)\,{\oplus}\,q(v)\,{=}\,(u_iv_j\,{\oplus}\,u_jv_i)_{1\,{\leq}\,i\,{<}\,j\,{\leq}\,n}.
\end{equation}
By \eq{phasepair}, Phase gates on the incoming EXORs $u,v$ and the newly computed EXOR $u\,{\oplus}\,v$ supply $\czgate(u,v)$.  Its coefficient vector is exactly \eq{increment}.  For disjoint supports this is the rectangle in \eq{rectangle}; shared variables contribute linear corrections, while repeated pairs cancel.  The expression is symmetric in $u,v$ and has zero $\czgate$ vector when $u\,{=}\,v$.  Chronological increments and the newly available target $\czgate$ vectors span the same space, by \eq{increment}.  This gives an equivalent rank calculation when every new parity has a Phase slot, including the packed checkerboard schedules below.

\subsection{The depth lower bound and Phase packing}\label{ssec:bound}

\begin{lemma}[Lower bound]\label{lem:depthbound}
For $n\,{\geq}\,2$, a universal programmable $\cnotgate{+}\pgate$ circuit satisfying \eq{fullrank} requires at least $D(n)$ $\cnotgate$ gates and $\cnotgate$ depth at least
\begin{equation}\label{eq:depthbound}
d\,{\geq}\,\left\lceil\frac{\binom{n}{2}}{\lfloor n/2\rfloor}\right\rceil
\,{=}\,\begin{cases}n,&n\text{ odd},\\n{-}1,&n\text{ even}.\end{cases}
\end{equation}
This bound holds even over unrestricted connectivity.
\end{lemma}

\begin{proof}
The initial $\czgate$ vectors $q(e_1),q(e_2),...,q(e_n)$ are zero.  By \eq{wireupdate}, a $\cnotgate$ changes only one wire parity and therefore can add at most one basis vector to the available $\czgate$ span.  Spanning all $D(n)$ $\czgate$ dimensions requires at least $D(n)$ $\cnotgate$ gates.  A matching layer contains at most $\lfloor n/2\rfloor$ $\cnotgate$ gates, giving \eq{depthbound}.
\end{proof}

We note that the result of this Lemma generalizes to ancilla-free universal programmable Clifford circuits composed with $\cnotgate, \pgate,$ and $\hgate$ gates, with a constant number of Hadamard stages to offer the lower bound of $4n\,{+}\,O(1)$.  This has particular value in light of the upper bound in \lem{unitary}.

Define $d_0(n):=2\lceil\frac{n}{2}\rceil \,{-}\,1$, the counting lower bound in \eq{depthbound}.  At equality in \eq{depthbound}, the number of available gate positions is exactly $D(n)$.  Thus, every layer must contain $K\,{:=}\,\lfloor n/2\rfloor$ gates, and every gate must add one new CZ basis vector.  These observations provide strong exact pruning rules for the searches in \sec{search}.  When an additional layer is allowed, as needed at $n\,{=}\,4$ and $n\,{=}\,8$, dependent intermediate gates must also be permitted.

Most successful circuits alternate the checkerboard color of the controls.  In such a circuit, a target of odd-numbered layer $2j{-}1$ is a control or an idle wire in layer $2j$.  A Phase gate on this wire commutes through that next layer.  Therefore, Phase insertions after every $\cnotgate$ layer can be consolidated into rounds after layers
$
2,4,...,2\lfloor d/2\rfloor,
$
with an additional round after layer $d$ (when $d$ is odd---and with a few exceptions it is).  Phase gates are also applied to the original inputs.  This consolidation calls for a total of $\lceil d/2\rceil{+}1$ Phase gate rounds.

\subsection{A worst-case state depth bound for arbitrary Clifford circuits in a lattice}\label{ssec:statebound}

\lem{depthbound} concerns universally programmable circuits  of the form -$\hgate$-\{$\cnotgate$+$\pgate$\}-$\lcgate$-.  We next give a lower bound that allows the entangling circuit and the local Clifford gates to depend on the desired state, and does not require a fixed entangling schedule.  Obviously, such unrestricted circuits offer more freedom, and the associated lower bound will be smaller than that in \lem{depthbound}.

\begin{lemma}[Clifford state lower bound]\label{lem:statebound}
Consider a fixed labeled $n$-qubit patch of the square lattice, with arbitrary single-qubit Clifford gates and layers of disjoint nearest-neighbor $\cnotgate$, controlled-Y ($\cygate$), and $\czgate$ gates, without ancillary qubits or measurements.  If depth $d$ suffices to prepare every pure $n$-qubit Clifford state, then
\begin{align}\label{eq:statebound}
d&\,{>}\,\frac{n{+}3{-}2\log_2(6)}{2\log_2\!\left((1{+}\sqrt{73})/2\right)}\,{=}\,0.221769358\ldots\,n{-}0.481222875\ldots.
\end{align}
\end{lemma}

\begin{proof}
The number of pure $n$-qubit stabilizer states, modulo global phase, is \cite[Proposition 2]{aaronson2004improved}
\begin{equation}\label{eq:statecount}
S_n\,{=}\,2^n\prod_{j\,{=}\,1}^n(2^j{+}1),\qquad
\log_2(S_n)\,{>}\,\frac{n(n{+}3)}{2}.
\end{equation}
We bound the number of states obtainable in depth at most $d$ by first removing redundant single-qubit choices.

Since $\cnotgate(c,t)\,{=}\,\hgate(t)\czgate(c,t)\hgate(t)$ and $\cygate(c,t)\,{=}\,\pgate(t)\cnotgate(c,t)\pgate^\dagger(t)$, any matching layer of $\cnotgate$, controlled-Y, and $\czgate$ gates can be replaced by a $\czgate$ matching and local Clifford gates at no additional two-qubit depth.  Write the resulting circuit as
\begin{equation}\label{eq:statenormalform}
U\,{=}\,C_dE_dC_{d{-}1}E_{d{-}1}...C_1E_1C_0,
\end{equation}
where $E_j$ is a $\czgate$ matching and $C_j$ is local Clifford.

Modulo global phase, the single-qubit Clifford group has $24$ elements.  Quotienting by its four-element normal Pauli subgroup leaves six classes; the subgroup generated by $\pgate$ has two classes in this quotient.  Thus, every single-qubit Clifford can be written as
\begin{equation}\label{eq:localnormalform}
C\,{=}\,R\pgate^a Q,\qquad
R\,{\in}\,\{\idgate,\hgate,\pgate\hgate\},\quad
a\,{\in}\,\{0,1\},\quad
Q\,{\in}\,\{\idgate,\xgate,\ygate,\zgate\}.
\end{equation}
Here, operator products act from right to left, as in \eq{statenormalform}.

Starting at $C_d$ and proceeding toward the input, apply \eq{localnormalform} on each qubit occupied by $E_j$.  The power of $\pgate$ commutes through $E_j$, and conjugating a Pauli by $E_j$ produces another product of single-qubit Paulis.  Both can therefore be absorbed into $C_{j{-}1}$.  On an idle qubit, the entire local Clifford commutes through $E_j$ and is absorbed into $C_{j{-}1}$.  Repeating this procedure leaves only one of the three representatives $R$ on each occupied qubit and no local choice on an idle qubit.  The remaining input layer $C_0$ prepares one of the six single-qubit stabilizer states on each wire, giving at most $6^n$ product inputs.

Count a normalized layer by deleting vertices.  In any nonempty residual patch, choose a vertex minimizing $x{+}y$.  It has no left or lower neighbor, hence at most two neighbors.  If it is idle, delete it.  Otherwise, choose one of at most two neighbors and one of $3{\times}3\,{=}\,9$ pairs of endpoint representatives, and delete both endpoints.  This argument remains valid after every deletion, even if the residual patch is disconnected or has holes.

It follows by induction that a patch with $m$ vertices has at most $F_m$ normalized layer descriptions, where
\begin{equation*}
F_0\,{=}\,F_1\,{=}\,1,\qquad
F_m\,{=}\,F_{m{-}1}{+}18F_{m{-}2}\quad(m\,{\geq}\,2).
\end{equation*}
Let $r\,{:=}\,(1{+}\sqrt{73})/2$.  Since $r^2\,{=}\,r{+}18$, induction gives $F_m\,{\leq}\,r^m$.  Padding shorter circuits with empty matchings, the number $N_d$ of states obtainable in depth at most $d$ satisfies
\begin{equation}\label{eq:normalizedstatecount}
N_d\,{\leq}\,6^nF_n^d\,{\leq}\,6^nr^{nd}.
\end{equation}
If every stabilizer state can be prepared, then $S_n\,{\leq}\,N_d$.  Combining \eq{statecount} and \eq{normalizedstatecount} gives
\begin{equation*}
\frac{n(n{+}3)}{2}\,{<}\,n\log_2(6){+}nd\log_2(r),
\end{equation*}
which rearranges to \eq{statebound}.
\end{proof}

Although the leading coefficient in \eq{statebound} is conservative, it is only a factor $2\log_2\!\left((1{+}\sqrt{73})/2\right)\,{=}\,4.509189\ldots$ below the coefficient one in the constructive depth $d_0(n)$ achieved by the fully programmable cycle and search constructions below.  The remaining gap is therefore only about a factor of $4.5$, despite the stronger requirement of full programmability and layered structure in our constructions.  Only square-lattice connectivity is used in the lower-bound argument.

The same count applies to Clifford unitary synthesis in the gate library of \lem{statebound}, with $24^n$ input local Clifford choices in place of $6^n$ product inputs.  Since there are $2^{2n^2{+}O(n)}$ Clifford unitaries modulo global phase, comparison with $24^nF_n^d$ gives the depth lower bound
\[
d\,{\geq}\,\frac{2}{\log_2\!\left((1{+}\sqrt{73})/2\right)}\,n{+}O(1)
\,{=}\,0.887077433\ldots\,n{+}O(1).
\]

\subsection{Matching obstructions on general connectivity graphs}\label{ssec:matching}

The rank criterion and \lem{depthbound} apply to any connectivity graph $G$, not only the square lattice.  We call a circuit satisfying \eq{fullrank} $\czgate$-universal, and write $d_0(n)$ for the right-hand side of \eq{depthbound}.  Small diameter might appear favorable: parities can spread across the graph quickly.  However, attaining $d_0(n)$ also imposes matching conditions that diameter alone does not capture.

\begin{lemma}[Matching obstructions]\label{lem:matching}
Let $G$ have $n\,{\geq}\,3$ vertices and admit a $\czgate$-universal $\cnotgate$ circuit of depth $d_0(n)$. Then:
\begin{enumerate}
\item Consecutive circuit layers are edge-disjoint maximum matchings.
\item For even $n$, $G$ contains two edge-disjoint perfect matchings.  The subgraph consisting of all edges that belong to a perfect matching is connected.  In particular, $G$ has no cut vertex, bridge, or leaf.
\item For odd $n$, $G$ contains two edge-disjoint matchings, each leaving exactly one vertex unmatched.  Their union consists of even cycles and one even-length path, possibly a single isolated vertex.  In particular, $G$ has at most two leaves.
\end{enumerate}
\end{lemma}

\begin{proof}
At depth $d_0(n)$, every gate must supply an independent $\czgate$ vector.  If consecutive layers reuse an edge with incoming parities $u,v$, its first use replaces them by $u,u\,{\oplus}\,v$, or $u\,{\oplus}\,v,v$.  Its second use therefore generates the linear functions already seen and does not offer a new CZ dimension.  Such a repeated edge is thus prohibited.

For even $n$, every layer is a perfect matching.  If the subgraph formed by all perfect-matching edges were disconnected, every circuit layer would preserve its components.  No parity, and hence no $\czgate$ vector, could then involve variables from different components.  To exclude a cut vertex $v$, remove it and consider the resulting components.  A perfect matching requires exactly one component of odd order: the component containing the vertex matched to $v$.  Edges from $v$ to any other component, of even order, belong to no perfect matching.  Thus the perfect-matching subgraph would be disconnected.  A connected graph with $n\,{>}\,2$ and a bridge has a cut vertex, and a leaf is incident to a bridge.

For odd $n$, the union of two edge-disjoint maximum matchings has degree two except at their unmatched vertices.  Alternating edges give even cycles and an even-length path between the unmatched vertices; if those vertices coincide, the path has length zero.  A leaf must be unmatched in at least one of the two layers, since otherwise its only edge would be repeated.  There are only two unmatched positions.
\end{proof}

There is a useful consequence for a graph of degree three.  Two edge-disjoint perfect matchings leave a third perfect matching, so a necessary condition is that the edges can be colored with three colors, each color forming a matching.  The Petersen graph, of diameter two, fails this condition: removing any perfect matching leaves two disjoint five-cycles, which cannot contain a second perfect matching.  The same obstruction applies to every cubic graph that cannot be three-edge-colored \cite{karabas2024cubic}.  Thus, even a very small diameter and the absence of leaves do not suffice to guarantee the depth optimality we are seeking here.

\subsection{Linear nearest-neighbor connectivity and trees}\label{ssec:trees}

\begin{lemma}[Paths and trees]\label{lem:trees}
No tree on $n\,{\geq}\,4$ vertices admits a $\czgate$-universal $\cnotgate$ circuit of depth $d_0(n)$.  In particular, LNN connectivity cannot attain this depth. 
\end{lemma}

\begin{proof}
For even $n$, \lem{matching} excludes every tree.  For odd $n\,{=}\,2m{+}1$, two edge-disjoint maximum matchings use $2m\,{=}\,n{-}1$ edges.  In a tree they must use every edge, giving degree at most two.  Thus the tree must be the path $1,2,...,2m{+}1$.

On this path, the only possible consecutive edge-disjoint maximum matchings are, up to reflection,
\begin{equation*}
A\,{=}\,\{(1,2),(3,4),...,(2m{-}1,2m)\},\qquad
B\,{=}\,\{(2,3),(4,5),...,(2m,2m{+}1)\}.
\end{equation*}
They must alternate.  Before the third layer, each pair of $A$ must have received a target in the second layer, or its increment would repeat.  There are $m$ targets and $m$ such pairs.  The left boundary therefore forces $(2,3)$ to target vertex $2$, then $(4,5)$ to target vertex $4$, and so on.  All second-layer targets are even.  Applying the same reasoning to the pairs of $B$ forces all third-layer targets to be even as well.

Fix an even vertex, and let the parities on its left neighbor, itself, and its right neighbor just after layer one be $a,b,c$.  The neighbors remain unchanged in layers two and three.  By \eq{increment}, the four increments at this vertex's incident gates cancel in the $\czgate$ coefficient space:
\begin{equation*}
\begin{aligned}
&\czgate(a,b)\,\,\czgate(b,c)\,\,\czgate(a,b\,{\oplus}\,c)
\,\,\czgate(c,a\,{\oplus}\,b\,{\oplus}\,c) \sim \idgate.
\end{aligned}
\end{equation*}
Expanding the two EXOR arguments leaves each pair twice, while $\czgate(c,c)$ has zero $\czgate$ vector.  The first increment may be evaluated after its gate, since adding the control to the target preserves its $\czgate$ vector.  The fourth increment is unchanged by reversing that gate's direction.  It is therefore dependent, contradicting the requirement that every gate be productive.  For $n\,{\geq}\,5$, the required fourth layer exists.

Observe that one gate suffices on two vertices, and the three chronological gates $\cnotgate(1,2)$, $\cnotgate(3,2)$, and $\cnotgate(1,2)$ span all three pair coordinates on the three-vertex path, explaining the selection $n\,{\geq}\,4$ in the statement of Lemma.
\end{proof}

In the next subsection, we consider cycles and show that, in contrast to the LNN, the minimal depth can be reached.  The difference between LNN and a cycle is only in one edge, yet it turns out to be the key to enabling the minimal-depth construction.  Also, the known bound, $2n{+}2$ over the LNN (reducible to $2n{-}2$ using the 4-stage removal technique of \cite{maslov2023cnot}), is reduced by our new construction to $d_0(n)$, i.e., by a factor of about 2, being an impressive feat for a single edge to accomplish.

\subsection{Cycles}\label{ssec:cycles}

Cycles provide a positive result despite having a large diameter.  Number the vertices of an $n$-vertex cycle graph $C_n$ by $0,1,2,...,n{-}1$, cyclically, and let $m\,{=}\,\lfloor n/2\rfloor$ and $\epsilon\,{=}\,n\bmod2$.  At step $t\,{=}\,0,1,2,...$, use the layer
\begin{equation}\label{eq:cyclelayers}
\calL_t\,{=}\,\{\cnotgate(t{+}2j{+}\epsilon,\ t{+}2j{+}\epsilon{+}1):j\,{=}\,0,1,2,...,m{-}1\},
\end{equation}
where vertex indices are reduced modulo $n$.  For even $n$, this simply alternates
\begin{equation*}
\begin{aligned}
L_1&\,{=}\,\cnotgate(0,1)\cdot\cnotgate(2,3)\cdots\cnotgate(n{-}2,n{-}1)\text{ and }\\
L_2&\,{=}\,\cnotgate(n{-}1,0)\cdot\cnotgate(1,2)\cdots\cnotgate(n{-}3,n{-}2).
\end{aligned}
\end{equation*}
For odd $n$, the idle vertex advances by one position per layer.  Phase gate applications after every second layer, with the final layer included, together with input Phase corrections, are sufficient to implement all CZ dimensions.  For final depth $d$, use cuts $1,3,5,...,d$ when $d$ is odd and $2,4,6,...,d$ when $d$ is even.

For example, on six qubits numbered $0,1,2,3,4,5$, execute $L_1,L_2,L_1,L_2,L_1$.  The $\czgate$ vectors available cumulatively at the Phase cuts after layers $1,3,5$ have ranks $3,9,15$, correspondingly.  The last rank equals $D(6)\,{=}\,\binom{6}{2}$, so five layers supply the full $\czgate$ space.  To see a particular program, observe that after layer three, wires $2$ and $3$ contain $x_0\,{\oplus}\,x_1\,{\oplus}\,x_2$ and $x_0\,{\oplus}\,x_1\,{\oplus}\,x_3$, correspondingly.  Apply $\pgate$ to wire $2$ and $\pgate^\dagger$ to wire $3$ at this cut, together with input corrections $\pgate^\dagger$ on wire $2$ and $\pgate$ on wire $3$ at the input boundary of the $\cnotgate$ circuit.  Leave all other Phase slots unused.  The resulting phase exponent satisfies
\begin{equation*}
\begin{aligned}
&(x_0\,{\oplus}\,x_1\,{\oplus}\,x_2){-}(x_0\,{\oplus}\,x_1\,{\oplus}\,x_3){-}x_2{+}x_3 \equiv 2x_0x_2{+}2x_0x_3{+}2x_1x_2{+}2x_1x_3\pmod4.
\end{aligned}
\end{equation*}
Thus, the programmed circuit implements $U_A\czgate(0,2)\czgate(0,3)\czgate(1,2)\czgate(1,3)$, where $A$ is the fixed linear reversible endpoint of the five layers.  Several of $\czgate$ gates are not cycle edges: the walk supplies their phases through intermediate parities.  

\begin{lemma}[Cycles]\label{lem:cycles}
For $n\,{\geq}\,3$, the minimum $\czgate$-universal $\cnotgate$ depth on the simple cycle $C_n$ is
\begin{equation}
d_C(n)\,{=}\,\begin{cases}
d_0(n),&n\,{=}\,4k{+}1,4k{+}2,4k{+}3,\\
d_0(n){+}1,&n\,{=}\,4k.
\end{cases}\nonumber
\end{equation}
The first $d_C(n)$ layers of \eq{cyclelayers} attain this depth.
\end{lemma}

\begin{proof}
The direct rectangle cancellations in \app{cycles} isolate all $D(n)$ standard $\czgate$ basis vectors after $n$ layers for odd $n$, after $n{-}1$ layers for $n\,{=}\,4k{+}2$, and after $n$ layers for $n\,{=}\,4k$.  Here we prove that the extra layer in the last case is necessary for any circuit on the cycle, allowing arbitrary gate directions.

Let $n\,{=}\,2m$ with even $m$, and suppose depth $n{-}1$ suffices.  Every layer is a perfect matching, and the two perfect matchings of the cycle must alternate by \lem{matching}.  Each pair in one layer must receive a target in the next layer, or its reuse in the following layer would repeat an increment.  Counting targets forces exactly one target per preceding pair.  Around a cycle, this forces each interior layer to point uniformly clockwise or uniformly counterclockwise.

The first layer can also be made uniform by an invertible change of input variables within each of its disjoint pairs: choose the change so that the two output rows equal those of a uniformly directed first layer.  Every invertible two-variable binary change preserves the pair's increment in \eq{increment}, while the common invertible change applied to all later increments preserves their rank.  The final layer's directions do not affect its incoming pair-coefficient vectors.  Consequently, within each layer its $m$ increments are rotations by two vertices of one vector.

Sum the increments of each layer.  Under rotation by two vertices there are $m$ orbits of cross-checkerboard pairs.  Within each checkerboard class, the orbit of opposite pairs has length $m/2$, so its contribution to a sum of $m$ rotations is zero over $\FF_2$.  The other same-class pairs form $m/2{-}1$ orbits per class.  All layer sums therefore lie in a space of dimension at most
\begin{equation*}
m{+}2(m/2{-}1)\,{=}\,n{-}2.
\end{equation*}
The $n{-}1$ layer sums are dependent.  This gives a nonempty sum of gate increments equal to zero, contradicting the independence of all $D(n)$ gates.  The other cases are optimal by \lem{depthbound}.
\end{proof}

\subsection{Hamiltonian minimum-perimeter square-lattice patches}\label{ssec:patches}

An odd cycle cannot be embedded in the square lattice.  This is because every left step in the directed cycle path must be matched by a right step, and every up step must be matched by a down step; thus, the cycle length is the sum of two even numbers and cannot be odd.  For even sizes, one must also choose a patch that supports a cycle through every vertex.  The following construction satisfies the geometric requirements simultaneously.

For an $n$-site patch, let $p\,{=}\,4n{-}2e$ be its edge perimeter, where $e$ counts physical adjacency edges.  A patch in an $h{\times}w$ bounding rectangle has $p\,{\geq}\,2(h{+}w)$ and $hw\,{\geq}\,n$, giving
\begin{equation}\label{eq:perimeter}
p\,{\geq}\,2\lceil2\sqrt n\rceil.
\end{equation}
Here, perimeter counts exposed unit sides of the occupied lattice cells.

\begin{lemma}[Minimum-perimeter Hamiltonian patches]\label{lem:patches}
For every even $n\,{\geq}\,4$, there is a connected, hole-free $n$-vertex square-lattice patch of perimeter $2\lceil2\sqrt n\rceil$ containing a Hamiltonian cycle.  Both dimensions of its bounding rectangle are $\sqrt n{+}O(1)$.
\end{lemma}

\begin{figure}[t]
\centering
\includegraphics[width=0.8\textwidth]{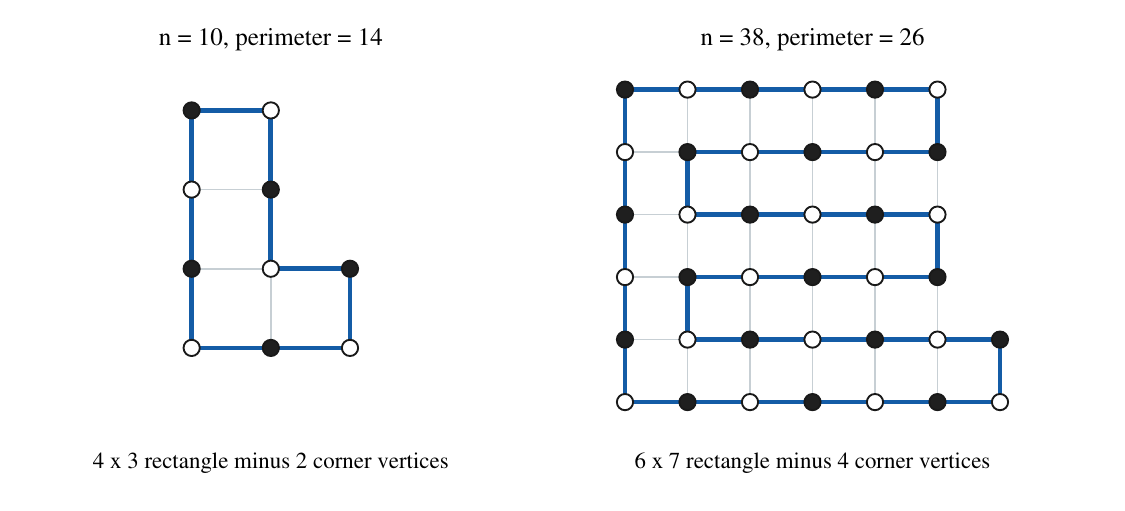}
\caption{Minimum-perimeter patches with Hamiltonian cycles for $n\,{=}\,10$ (left) and $n\,{=}\,38$ (right).  They are a $4{\times}3$ rectangle short of two corner vertices and a $6{\times}7$ rectangle short of four corner vertices, correspondingly.  Gray edges show physical adjacency, and blue edges form the cycle.}
\label{fig:patches}
\end{figure}

\begin{proof}
Set
\begin{equation}\label{eq:patchparameters}
s\,{=}\,\lceil2\sqrt n\rceil,\qquad
h\,{=}\,2\lfloor(s{+}1)/4\rfloor,\qquad w\,{=}\,s{-}h,\qquad r\,{=}\,hw{-}n.
\end{equation}
These choices give even $h$, $h,w\,{\geq}\,2$, and even $r$ with $0\,{\leq}\,r\,{\leq}\,h{-}2$; also $w\,{\geq}\,3$ whenever $r\,{>}\,0$.  To check the last bounds, write $s\,{=}\,4a,4a{+}1,4a{+}2,$ or $4a{+}3$.  The products $hw$ are, correspondingly, $4a^2$, $4a^2{+}2a$, $4a^2{+}4a$, and $4a^2{+}6a{+}2$.  Combining $(s{-}1)^2/4\,{<}\,n\,{\leq}\,s^2/4$ with even $n$ gives $0\,{\leq}\,hw{-}n\,{\leq}\,h{-}2$ in each case.

Take an $h{\times}w$ rectangle and remove the top $r$ vertices of its rightmost column.  This corner notch leaves the patch connected and hole-free, preserves its perimeter $2(h{+}w)\,{=}\,2s$, and removes equal numbers of black and white vertices.  The parameter bounds ensure that the last column retains at least two vertices.

Divide the patch into $h/2$ horizontal two-row strips.  Each strip is a rectangle of width $w$ or $w{-}1$, at least two, and has a cycle around its perimeter containing all its vertices.  To join two neighboring strip cycles, remove their two facing horizontal edges in the first two columns and insert the two vertical edges between them.  This replaces two cycles by one.  Repeating at all strip boundaries gives a Hamiltonian cycle, as illustrated in \fig{patches}.  Finally, \eq{patchparameters} gives $h,w\,{=}\,\sqrt n{+}O(1)$, and \eq{perimeter} proves minimum perimeter.
\end{proof}

\begin{cor}\label{cor:patchstates}
Every even $n\,{\geq}\,4$ admits a universal programmable Clifford state preparation circuit on a hole-free minimum-perimeter 2DSL patch, of depth at most $n{-}1$ for $n\,{=}\,4k{+}2$ and at most $n$ for $n\,{=}\,4k$.  The former depths attain \eq{depthbound}.
\end{cor}

\begin{proof}
Apply \lem{cycles} to the Hamiltonian cycle in \lem{patches}, and use \eq{endpointabsorb}.
\end{proof}

\section{Search algorithms}\label{sec:search}

The cycle construction settles the $n\,{=}\,4k{+}2$ family analytically, but leaves roughly three quarters of $n$ unaddressed.  This section settles the remaining values $n$ via a search, focusing on compact, hole-free, preferably minimum-perimeter layouts.  A reader may skip to \sec{unitaries} that extends the $n\,{=}\,4k{+}2$ family result to implement a Clifford unitary over 2DSL in depth $4n\,{+}\,O(1)$.

All searches use the same mathematical objective: find a 2DSL-local $\cnotgate$ circuit whose $\czgate$ vectors at the declared Phase slots span all $D(n)$ $\czgate$ dimensions.  At the counting bound, it is useful to score a circuit by the CZ dimension generated by its gates before the first dependent vector.  This independent-prefix score is cheaper to evaluate than the total rank of every failed circuit.  A full score of $D(n)$ certifies the same algebraic condition as \eq{fullrank}.

\subsection{Exact reductions and geometric choices}\label{ssec:reductions}

Several restrictions are necessary at equality in \eq{depthbound}.  Every layer is a maximum matching, and every gate contributes a new basis vector.  The physical layout must be connected, since $\cnotgate$s cannot produce a $\czgate$ interaction involving inputs from different connected components.  The smaller checkerboard class must have at least $K\,{=}\,\lfloor n/2\rfloor$ vertices.  A prescribed partial matching must admit a completion covering this class; a maximum bipartite matching calculation, or Hall's condition \cite{hall1935representatives}, can reject a geometric choice before a large rank computation. \lem{matching} further excludes repeated undirected edges in consecutive layers at the counting bound.  

The remaining geometric restrictions are choices of search family.  Alternating checkerboard directions, concentric first matchings, short periods, compact layouts, and preferences transferred from a parent square do not follow from the lower bound.  However, these heuristic choices help to limit the search space, since it grows very quickly. 

For nonsquare cases, we use connected compact patches, typically rectangles or rectangles with corner cells removed.  The secondary metric is the edge perimeter; see \eq{perimeter}.  Layout catalogs favor compact patches defined as hole-free, minimum-perimeter, containing two edge-disjoint maximum matchings.  

We first consider the small-case exclusions, $n\,{=}\,4$ and $n\,{=}\,8$.

\subsection{\texorpdfstring{Small exhaustive search, including $n\,{=}\,4$ and $n\,{=}\,8$}{Small exhaustive search, including n=4 and n=8}}\label{ssec:small}

The exact search generates connected square-lattice shapes by adding a boundary cell to every smaller shape.  Translation, rotation, and reflection identify equivalent layouts.  For each layout, enumerate every maximum matching and every orientation of its edges.  A recursive state consists of the current wire parities, an exact binary elimination basis, the preceding matching, and the layer number.  At the counting bound, reject a matching as soon as one of its new target $\czgate$ vectors is dependent.  Otherwise, branch over its $\cnotgate$ orientations and continue.  All rank insertions are undone when returning to the parent state.  A negative answer is valid only after the complete finite search tree has been exhausted.

For $n\,{=}\,4$, the counting bound is three.  There are five free connected four-cell shapes, and only the $2{\times}2$ square has two edge-disjoint perfect matchings.  On this square there are two perfect matchings and four orientations of each, giving $8^3\,{=}\,512$ dense depth-three circuits before any pruning.  Direct enumeration gives maximum $\czgate$ rank five, below $D(4)\,{=}\,6$.  A depth-four circuit with rank six is given in \app{small}.

For $n\,{=}\,8$, the counting bound is seven.  There are $369$ free connected eight-cell shapes, obtained from $2{,}725$ translation-distinct shapes.  Only four shapes have a pair of edge-disjoint perfect matchings.  The exact search over these layouts, their matching sequences, and all $\cnotgate$ directions has $28{,}916$ recursive states and $33{,}519$ matching-rank tests.  No full-rank depth-seven circuit exists.  The enumeration is independently reproduced with edge-subset matching generation and a separate elimination implementation.  A full-rank depth-eight circuit on a $2{\times}4$ rectangle supplies the matching upper bound.

For other small cases, finding a circuit satisfying \eq{depthbound} already proves depth optimality, even when the layout catalog or search order is restricted.  The depth-seven circuit for $n\,{=}\,7$ uses mixed checkerboard directions and is certified with Phase rounds after every layer.  At $n\,{=}\,16$, a depth-fifteen solution is obtained without imposing the rigid periodic form used by Algorithm A.  These cases motivate keeping a general small-circuit search alongside the structured algorithms discussed next.

\subsection{Algorithm L: \lem{cycles}/\lem{patches}, the cycle construction}\label{ssec:L}

For $n\,{=}\,4k{+}2$, Algorithm L is the formula in \eq{cyclelayers}, embedded along the Hamiltonian cycle of \lem{patches} and truncated after $n{-}1$ layers.  Its full span follows from \lem{cycles}; there is no search over directions, matchings, or Phase schedules.  We retain L in the computational table to record independently checked instances and their verification costs alongside the search results.  Its Total search time is zero.

\subsection{Algorithm A: A perfect square}\label{ssec:A}

Let $n$ be a perfect square, let $R$ rotate an entire directed matching clockwise by $90$ degrees, and put
\begin{equation}\label{eq:ABCB}
F\,{=}\,(A,B,C,B).
\end{equation}
The two occurrences of $B$ are identical.  Layers $A$ and $C$ have one control color and $B$ has the other.  For odd $n$, use the sixteen-layer period
\begin{equation}\label{eq:Aodd}
\Pi_{\rm odd}\,{=}\,(F,R(F),R^2(F),R^3(F)).
\end{equation}
For even $n$, use the eight-layer period
\begin{equation}
\Pi_{\rm even}\,{=}\,(F,R^2(F)).\nonumber
\end{equation}
Repeat the complete period and truncate at depth $n$ or $n{-}1$, correspondingly.  These rotations preserve the relevant checkerboard colors.  Consecutive-edge exclusions are enforced within a block, between rotated blocks, and across the period boundary.

\begin{figure}[t]
\centering
\includegraphics[width=0.88\textwidth]{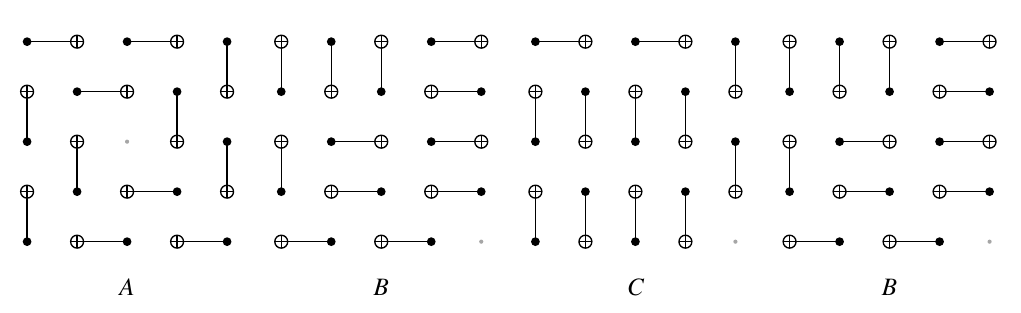}
\caption{The generating block $F\,{=}\,(A,B,C,B)$ of a certified depth-$25$ circuit on a $5\,{\times}\,5$ square.  Filled dots are $\cnotgate$ controls and circled plus signs are targets.  Small gray dots indicate idle sites.  The full circuit repeats the rotated sixteen-layer period of \eq{Aodd}, truncates it after layer $25$, and has $\czgate$ rank $300$.  A mutation of $B$ changes both displayed occurrences and all their rotated copies.}
\label{fig:ABCB}
\end{figure}

The main variant fixes $A$ to a concentric matching.  This was a largely arbitrary decision, but one that looked aesthetically appealing. Enumerate the vertices on each square ring clockwise, starting at its upper-left corner, and pair consecutive vertices.  The central vertex is idle when $n$ is odd.  Orient the pairs by the prescribed control color.  This $A$ is invariant under a quarter-turn for odd $n$ and a half-turn for even $n$.  Only $B$ and $C$ remain to be searched.  \fig{ABCB} shows one certified solution for $n\,{=}\,25$.

The optimizer keeps a population of matching triples.  It favors parents with a longer independent prefix, while retaining occasional lower-scoring candidates and fresh random starts.  Local mutations flip two parallel dominoes in a $2{\times}2$ cell to the other perfect matching of that cell.  Larger mutations remove several matching entries and repair them using randomized augmenting paths.  All occurrences of a generating matching change together.  Memoization avoids repeated evaluations of the same triple.  The implementations use a population of up to $96$ triples.

An alternative fixes $B$ to a shifted checkerboard tiling of horizontal and vertical domino cells and searches $A$ and $C$.  Exposed boundaries are completed by legal dominoes.  This floor-pattern variant supplied useful even-square solutions during the development of the search.  It uses the same periodic and rank framework; a half-turn must rotate the searched $A$ as well.  Neither variant is an arbitrary-side existence proof.

\subsection{Algorithm P: Periodic transfer to rectangles}\label{ssec:P}

Algorithm P retains the four-layer block \eq{ABCB} on a rectangular layout.  The main families are $k\,{\times}\,(k{+}1)$ and $k\,{\times}\,(k{+}2)$ rectangles, with $n\,{=}\,k(k{+}1)$ and $n\,{=}\,k(k{+}2)$.  Both families attain the perimeter bound \eq{perimeter}.

Let $\rho$ be a half-turn of the rectangle.  Define $\widehat\rho$ to apply this half-turn and, if it swaps the checkerboard classes, reverse every $\cnotgate$ arrow.  Use the eight-layer period
\begin{equation}
\Pi_P\,{=}\,(F,\widehat\rho(F)),\nonumber
\end{equation}
repeated and truncated at the counting depth.  For a $k\,{\times}\,(k{+}1)$ rectangle, the half-turn swaps colors, so the arrow correction is required.  For a $k\,{\times}\,(k{+}2)$ rectangle, it preserves colors.  Reversing these arrows enforces the layer-color convention.

All three generating matchings $A,B,$ and $C$ may vary.  Partial preferences can be obtained by rotating, reflecting, and cropping matching patterns of independently certified covering squares, followed by repair of the rectangle boundary.  The transferred object is a short-period pattern and a collection of preferences.  Deleting vertices from a certified square circuit does not itself certify the smaller circuit.  In particular, fixing the square's concentric $A$ on a rectangle can obstruct completion at a corner; allowing $A$ to change avoids imposing that restriction.  The optimizer otherwise uses the same population, domino flips, matching repair, and exact prefix objective as Algorithm A.

\subsection{Gmg: Guided matching growth}\label{ssec:Gmg}

Gmg constructs a circuit one productive matching layer at a time.  It imposes no fixed first or final layer and no eight- or sixteen-layer period.  Each retained prefix stores its circuit, wire parities, and exact elimination state.  A matching-extension step proceeds as follows:
\begin{enumerate}
\item Remove edges used in the preceding layer.  For every remaining physical edge, compute the $\czgate$ vector of the prospective new target parity and reduce it against the current prefix span.
\item Choose an unmatched vertex in the smaller checkerboard class with the fewest admissible neighbors.  Try those neighbors in a randomized, guided order.
\item Tentatively insert the edge's $\czgate$ vector into the temporary basis.  Reject a dependent vector, a vertex collision, or a partial matching that cannot be completed geometrically.  Otherwise, recurse on the remaining vertices.
\item On completing a $K$-edge matching, update the wire parities and retain the extended independent prefix.  On backtracking, undo all temporary pivots and edge choices.
\end{enumerate}
At the counting bound these rank tests are exact consequences of \lem{depthbound}.  The matching recursion has a finite node budget, so a budget exit means that an extension was not found, rather than that none exists.

The search keeps pools of diverse prefixes, with up to $256$ retained prefixes at a given depth in the implemented family.  It preferentially extends deep prefixes, cuts back a few layers to obtain a different continuation, and occasionally restarts from the empty circuit.  Layout allocation is adaptive: promising layouts receive more visits, while exploration of other layouts continues.  A soft preference for extending paths in the union of consecutive matchings discourages premature closing of short cycles without forbidding them.

Guidance comes from independently certified larger square circuits.  Their layers are transferred under the eight symmetries of the square and cropped onto the chosen smaller layout.  The resulting preferred edges bias the matching order; all otherwise admissible edges remain available.  The original guided implementation uses such preferences in approximately $85\%$ of matching constructions.  Parents must be available before their guides are used, but each child is verified independently on its own geometry and phase schedule.

Two implementation improvements are useful for hard prefixes.  First, exact Hall or maximum-matching tests reject partial geometric matchings that cannot cover the remaining anchor vertices.  Second, moving between prefixes with a common ancestor can undo the discarded suffix and replay only the new suffix.  The $\cnotgate$ updates and the list of inserted pivots are reversible, so the restored state can be checked against full reconstruction.  Near completion, candidate $\czgate$ vectors may also be represented in the quotient by the fixed prefix span.  This requires a complete quotient reduction, or explicitly tracked coordinates; stopping at the first absent pivot is insufficient to define a canonical quotient representative.

\subsection{Periodic obstructions and exact rank evaluation}\label{ssec:obstructions}

The short periods of A and P permit algebraic rejection tests.  For incoming control and target parities $u,v$, the chronological gate increment is the $\czgate$ vector of $\czgate(u,v)$, by \eq{phasepair} and \eq{increment}.  These increments and the newly available target $\czgate$ vectors span the same space when all such parities are retained.  For checkerboard circuits, the packed Phase cuts preserve this span.  As in \app{cycles}, sums and ranks below refer to $\czgate$ coefficient vectors, with linear Phase corrections handled separately.

Let $M$ be the endpoint matrix of one period, and let $(u_1,v_1),(u_2,v_2),...,(u_g,v_g)$ be its incoming gate parities.  In period number $r$, beginning at $r\,{=}\,0$, the increments are
\begin{equation}\label{eq:periodorbit}
\czgate(u_jM^r,v_jM^r),\qquad j\,{=}\,1,2,...,g.
\end{equation}
The period acts by applying the same substitution to both arguments of each $\czgate$ gate.  Expanding the two EXORs gives all cross products, as in the rectangle identity \eq{rectangle}.  Repeated pairs cancel, while squares contribute only linear Phase corrections.  This substitution therefore defines a linear transformation of the $\czgate$ coefficient space.  Exact rank evaluation can be performed by replaying the parities and expanding the resulting $\czgate$ circuits.

\begin{prop}\label{propos:fixedspace}
Let $E\,{=}\,\ker(M\,{\oplus}\,I)$ have dimension $f$, and choose a basis $h_1,h_2,...,h_f$ of column vectors in $E$.  Put $\pi(u)\,{=}\,(uh_1,uh_2,...,uh_f)$.  If
\begin{equation}\label{eq:fixedspace}
\rank_{\FF_2}\{\czgate(\pi(u_j),\pi(v_j)):j\,{=}\,1,2,...,g\}
\,{<}\,\binom{f}{2},
\end{equation}
where the gate arguments are parities of the $f$ projected variables, then repeating this period, however many times, cannot produce a full $\czgate$ span.
\end{prop}

\begin{proof}
Since $M h_i\,{=}\,h_i$, we have $\pi(uM^r)\,{=}\,\pi(u)$ for every $r$.  Thus, all increments in \eq{periodorbit} project to the same list of $\czgate$ circuits in \eq{fixedspace}, independently of the period number.  The map $\pi$ is onto $\FF_2^f$, so each pair of projected variables has a pair of original parities mapping to it.  Expanding these original parities expresses their $\czgate$ circuit as a binary sum of original $\czgate$ basis vectors, up to linear Phase corrections.  Consequently, the projection of the full original $\czgate$ space contains every projected pair and has dimension $\binom{f}{2}$.  A deficient projected span therefore excludes a full original span.
\end{proof}

A rejection by \eq{fixedspace} has an explicit nonzero alternating-form witness.  More generally, let $B$ be a nonzero alternating bilinear form over $\FF_2$, meaning that $B(u,u)\,{=}\,0$ for every $u$.  Alternation ensures that $B(u,v)$ defines a linear functional of the $\czgate$ vector of $\czgate(u,v)$.  If $B(uM,vM)\,{=}\,B(u,v)$ for all $u,v$ and $B(u_j,v_j)\,{=}\,0$ for every gate of a period, this functional vanishes on every increment in every repeated period and excludes full span.  Such a witness can be checked by propagating it through one actual period.  Mutations that break a known witness are preferred over mutations preserving it.  Breaking one witness does not establish success; another obstruction or a finite-horizon dependency may remain.  The fixed-space dimension alone is not a rejection criterion.

The large rank computations remain exact over $\FF_2$.  $\czgate$ vectors are stored as packed bit vectors, small independent prefixes are screened first, and longer survivors are evaluated in successively larger rectangular rank problems.  Pivot order is chosen to recover the first dependent chronological vector, preserving the search objective.  Checkpoint continuation retains unfinished exact work.  At full scale, a dense $D(n){\times}D(n)$ matrix requires $\Theta(n^4)$ bits, which explains why rank evaluation becomes a bottleneck even when the generating circuit has a short description.

Earlier development searches evaluated whole circuits using simulated annealing, direction flips, matching rewires, layer replacements, and evolutionary crossover.  Near rank $D(n){-}1$, a null vector identified the missing hyperplane and guided local repairs.  These methods supplied useful small and intermediate examples.  The later A, P, and Gmg algorithms reduce the search space or reuse exact prefix information more aggressively.  

\section{Numeric results and verification}\label{sec:results}

\tab{allresults} lists computational results for $\ResultCount$ independently certified sizes: $\SmallCount$ from the Small search, $\ACount$ from A, $\PCount$ from P, $\GmgCount$ from Gmg, and $\LCount$ from the explicit construction L.  Every $n\,{=}\,4,5,...,\ContiguousN$ is represented, with additional sizes through $n\,{=}\,\LargestN$.  All circuits have $\czgate$ dimension $\binom{n}{2}$.  Of these, $\CountingCount$ attain the minimum depth formula \eq{depthbound}; $n\,{=}\,4,8$ attain the exact optima of 4 and 8.  

All selected layouts are connected and hole-free.  The $n\,{=}\,12$ layout has perimeter $16$ instead of the minimum $14$; every other layout attains \eq{perimeter}.  The $n\,{=}\,4,7$ circuits also relax checkerboard directions.  The general small search, rather than the periodic square algorithm A, supplies the $n\,{=}\,16$ example.  These distinctions are described in \tab{allresults} footnotes.

The campaign verifier was separate from the search program.  It checks physical adjacency and disjointness, then computes the wire parities and exact binary rank at the declared Phase gate insertions via Gaussian elimination.  A separate `Independent proof (s)' column reports its accumulated runtime. 

\begin{figure}[!ht]
\centering
\includegraphics[width=0.8\textwidth,trim=0 18 0 10,clip]{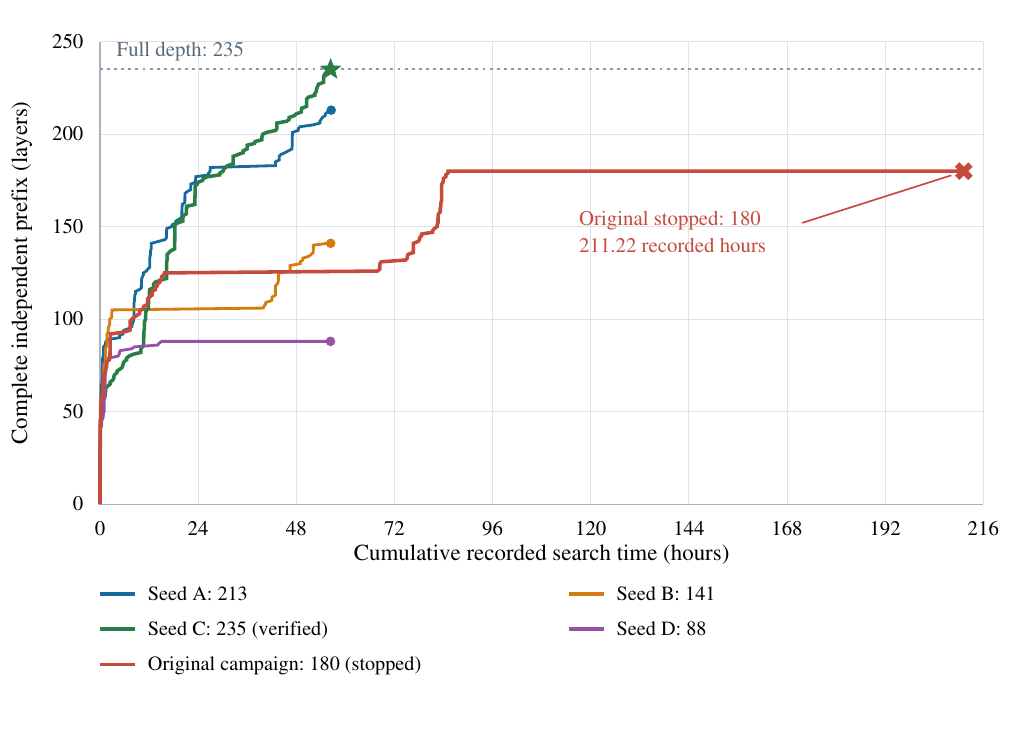}
\caption{Five searches for a depth-$235$ universal programmable $\cnotgate{+}\pgate$ circuit on $236$ qubits.  Each curve shows the evolution of the longest complete independent prefix with time.  The original Obsidian search (red) was stopped at $180$ layers after $211.22$ recorded hours.  One of four new independently seeded search threads reached depth $235$ after $56.39$ recorded hours.  These runs suggest that, when rerunning the search algorithm, it can be beneficial to launch multiple independent search threads rather than a single thread, since a single thread may get stuck.}
\label{fig:n236searches}

\vspace{4pt}
\includegraphics[width=0.8\textwidth,trim=0 18 0 10,clip]{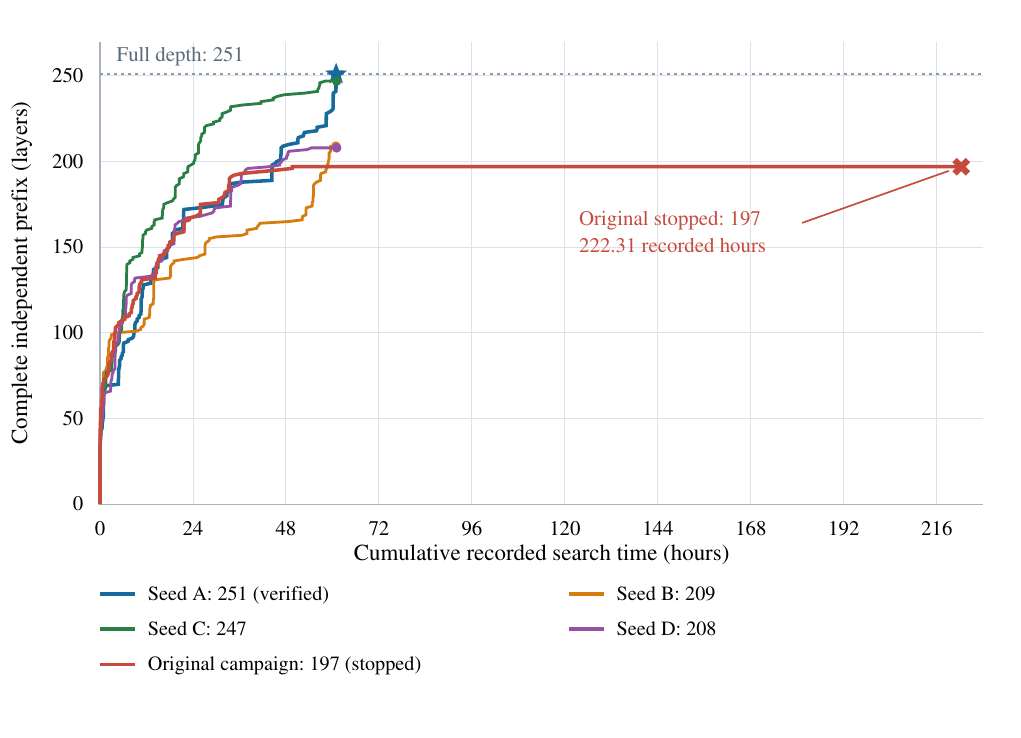}
\caption{Five searches for a depth-$251$ universal programmable $\cnotgate{+}\pgate$ circuit on $252$ qubits.  Each curve shows the evolution of the longest complete independent prefix against its own cumulative recorded search time.  The original Obsidian cluster search (red) was stopped at $197$ layers after $222.31$ recorded hours.  One of four new independently seeded search lineages reached depth $251$ after $60.95$ recorded hours.  Similarly to $n\,{=}\,236$, these runs show why independent parallel reruns can be useful when a single search gets stuck.}
\label{fig:n252searches}
\end{figure}

Search times sum the reported campaign's successful and unsuccessful search parts.  In particular, the failed P attempts at $n\,{=}\,20,24$ remain charged before Gmg succeeds.  Also, the five histories for $n\,{=}\,236$ and $n\,{=}\,252$ in \fig{n236searches} and \fig{n252searches}, correspondingly, illustrate the variability between independent reruns of Gmg. For L, there is no search time as the solution is constructed algorithmically.  The heterogeneous machines, different search spaces, and distinct timing scopes make the table a record of observed effort rather than a controlled speed comparison.

\clearpage
\begingroup
\footnotesize
\setlength{\tabcolsep}{1.5pt}
\setcounter{LTchunksize}{10}
\renewcommand{\arraystretch}{1.08}
\setlength{\LTcapwidth}{\textwidth}
\begin{longtable}{@{}rcrrrr||rcrrrr@{}}
\caption{All $\ResultCount$ independently certified universal programmable $\cnotgate{+}\pgate$ circuits explicitly computed.  $n$ counts the number of qubits; Algorithm identifies Small, A, P, or Gmg from \sec{search}, or the explicit cycle construction L; Depth counts 2DSL-compliant $\cnotgate$ gate layers; $\czgate$ Dimension confirms the equality to $\binom{n}{2}$.  Total search (s) sums timed search-engine wall seconds across successful and unsuccessful parts for that size.  Independent proof (s) reports the separate verifier's runtime.   The primary hosts were IonQ's Obsidian-8/9 cluster nodes, each with four $48$-core Intel Xeon Platinum 8468H processors ($192$ physical cores, $384$ logical CPUs), $385{,}262$ MiB RAM, Ubuntu 24.04 and Linux 6.8.  The expanded Obsidian-3--10 pool includes hosts with $128$ or $192$ physical cores, two hardware threads per core, and $385{,}262$--$2{,}321{,}813$ MiB RAM.  Each Small/Gmg worker uses one allocated physical core; A/P use one or four, with at most four threads for dense elimination.  No GPU was used.  Search and independent rank engines are written in C\#, built with .NET SDK 8.0.426 NativeAOT and clang 18.1.3 in Release/Speed mode with AVX2 and LZCNT; Python coordinates jobs, geometry, and small independent checks.  Footnotes identify various exceptions.}\label{tab:allresults}\\
\hline
$n$ & Algorithm & Depth & \shortstack{$\czgate$\\Dimension} & \shortstack{Total search\\(s)} & \shortstack{Independent\\proof (s)} & $n$ & Algorithm & Depth & \shortstack{$\czgate$\\Dimension} & \shortstack{Total search\\(s)} & \shortstack{Independent\\proof (s)} \\ \hline
\endfirsthead
\multicolumn{12}{c}{\tablename~\thetable\ (continued)}\\
\hline
$n$ & Algorithm & Depth & \shortstack{$\czgate$\\Dimension} & \shortstack{Total search\\(s)} & \shortstack{Independent\\proof (s)} & $n$ & Algorithm & Depth & \shortstack{$\czgate$\\Dimension} & \shortstack{Total search\\(s)} & \shortstack{Independent\\proof (s)} \\ \hline
\endhead
\hline
\multicolumn{12}{r}{Continued on next page}\\
\endfoot
\hline
\endlastfoot
$4$\footnote{Depth $4$ exceeds $n{-}1\,{=}\,3$, but is optimal by the exhaustive small-case analysis in \ssec{small}.  The circuit uses mixed checkerboard directions and Phase slots after every layer.} & Small & $4$ & $6$ & $0.00162$ & $0.00267$ & $5$ & Small & $5$ & $10$ & $0.00108$ & $0.00255$ \\
$6$\footnote{L is the explicit $4k{+}2$ cycle construction in \lem{cycles}, not a search.  Its search time is zero; deterministic generation and independent certification are separate.} & L & $5$ & $15$ & $0$ & $0.00064$ & $7$\footnote{The circuit uses mixed checkerboard directions and Phase slots after every layer; its depth $7$ attains \eq{depthbound}.} & Small & $7$ & $21$ & $0.00298$ & $0.00242$ \\
$8$\footnote{Depth $8$ exceeds $n{-}1\,{=}\,7$, but it is optimal by the exhaustive small-case analysis in \ssec{small}.} & Small & $8$ & $28$ & $0.00237$ & $0.00271$ & $9$ & A & $9$ & $36$ & $0.00088$ & $0.00233$ \\
$10$ & L & $9$ & $45$ & $0$ & $0.00112$ & $11$ & Small & $11$ & $55$ & $1.55611$ & $0.21676$ \\
$12$\footnote{The selected layout has perimeter $16$, whereas \eq{perimeter} gives $14$.  This is the only perimeter exception in the table.} & Small & $11$ & $66$ & $1.67816$ & $0.36806$ & $13$ & Small & $13$ & $78$ & $2.96250$ & $0.36571$ \\
$14$ & L & $13$ & $91$ & $0$ & $0.00125$ & $15$ & Small & $15$ & $105$ & $2.71429$ & $0.42298$ \\
$16$\footnote{The depth-$15$ circuit on the $4{\times}4$ square comes from the general small search, rather than Algorithm A.  It has minimum perimeter and alternating checkerboard directions.} & Small & $15$ & $120$ & $2.22729$ & $0.59543$ & $17$ & Gmg & $17$ & $136$ & $0.08170$ & $0.00320$ \\
$18$ & L & $17$ & $153$ & $0$ & $0.00169$ & $19$ & Gmg & $19$ & $171$ & $0.01804$ & $0.00355$ \\
$20$\footnote{Total search includes an unsuccessful Algorithm P attempt of approximately $10{,}800$ seconds before the successful Gmg attempt.} & Gmg & $19$ & $190$ & $10{,}800.00378$ & $0.00314$ & $21$ & Gmg & $21$ & $210$ & $0.06826$ & $0.00364$ \\
$22$ & L & $21$ & $231$ & $0$ & $0.00173$ & $23$ & Gmg & $23$ & $253$ & $0.05095$ & $0.00380$ \\
$24$\footnote{Total search includes an unsuccessful Algorithm P attempt of approximately $10{,}800$ seconds before the successful Gmg attempt.} & Gmg & $23$ & $276$ & $10{,}800.01910$ & $0.00337$ & $25$ & A & $25$ & $300$ & $1.54592$ & $0.50577$ \\
$26$ & L & $25$ & $325$ & $0$ & $0.00226$ & $27$ & Gmg & $27$ & $351$ & $0.08655$ & $0.00366$ \\
$28$ & Gmg & $27$ & $378$ & $0.06937$ & $0.00396$ & $29$ & Gmg & $29$ & $406$ & $0.14107$ & $0.00375$ \\
$30$ & L & $29$ & $435$ & $0$ & $0.00456$ & $31$ & Gmg & $31$ & $465$ & $0.84722$ & $0.00385$ \\
$32$ & Gmg & $31$ & $496$ & $0.08762$ & $0.00675$ & $33$ & Gmg & $33$ & $528$ & $0.23664$ & $0.00558$ \\
$34$ & L & $33$ & $561$ & $0$ & $0.00311$ & $35$ & P & $35$ & $595$ & $0.82490$ & $0.00564$ \\
$36$ & A & $35$ & $630$ & $1.57104$ & $0.42926$ & $37$ & Gmg & $37$ & $666$ & $1.11878$ & $0.00579$ \\
$38$ & L & $37$ & $703$ & $0$ & $0.00739$ & $39$ & Gmg & $39$ & $741$ & $0.24336$ & $0.00671$ \\
$40$ & Gmg & $39$ & $780$ & $1.35287$ & $0.00655$ & $41$ & Gmg & $41$ & $820$ & $0.32011$ & $0.00706$ \\
$42$ & L & $41$ & $861$ & $0$ & $0.00879$ & $43$ & Gmg & $43$ & $903$ & $4.47708$ & $0.00691$ \\
$44$ & Gmg & $43$ & $946$ & $4.06659$ & $0.00711$ & $45$ & Gmg & $45$ & $990$ & $1.04991$ & $0.00837$ \\
$46$ & L & $45$ & $1{,}035$ & $0$ & $0.00620$ & $47$ & Gmg & $47$ & $1{,}081$ & $2.69672$ & $0.00892$ \\
$48$ & P & $47$ & $1{,}128$ & $0.15126$ & $0.00956$ & $49$ & A & $49$ & $1{,}176$ & $1.59112$ & $0.45613$ \\
$50$ & L & $49$ & $1{,}225$ & $0$ & $0.00858$ & $51$ & Gmg & $51$ & $1{,}275$ & $6.04496$ & $0.00959$ \\
$52$ & Gmg & $51$ & $1{,}326$ & $14.33941$ & $0.01092$ & $53$ & Gmg & $53$ & $1{,}378$ & $2.32614$ & $0.01116$ \\
$54$ & L & $53$ & $1{,}431$ & $0$ & $0.01050$ & $55$ & Gmg & $55$ & $1{,}485$ & $0.83422$ & $0.01198$ \\
$56$ & P & $55$ & $1{,}540$ & $3.28045$ & $0.01268$ & $57$ & Gmg & $57$ & $1{,}596$ & $26.97884$ & $0.01377$ \\
$58$ & L & $57$ & $1{,}653$ & $0$ & $0.01325$ & $59$ & Gmg & $59$ & $1{,}711$ & $28.84386$ & $0.01452$ \\
$60$ & Gmg & $59$ & $1{,}770$ & $26.05394$ & $0.01455$ & $61$ & Gmg & $61$ & $1{,}830$ & $4.40011$ & $0.01371$ \\
$62$ & L & $61$ & $1{,}891$ & $0$ & $0.01574$ & $63$ & P & $63$ & $1{,}953$ & $0.51885$ & $0.01593$ \\
$64$ & A & $63$ & $2{,}016$ & $1.72594$ & $0.55039$ & $65$ & Gmg & $65$ & $2{,}080$ & $34.18852$ & $0.01825$ \\
$66$ & L & $65$ & $2{,}145$ & $0$ & $0.01885$ & $67$ & Gmg & $67$ & $2{,}211$ & $27.50502$ & $0.01843$ \\
$68$ & Gmg & $67$ & $2{,}278$ & $30.66003$ & $0.01941$ & $69$ & Gmg & $69$ & $2{,}346$ & $14.04888$ & $0.02003$ \\
$70$ & L & $69$ & $2{,}415$ & $0$ & $0.02247$ & $71$ & Gmg & $71$ & $2{,}485$ & $1.75185$ & $0.02162$ \\
\noalign{\penalty-10000}
$72$ & P & $71$ & $2{,}556$ & $1.10885$ & $0.02206$ & $73$ & Gmg & $73$ & $2{,}628$ & $27.73759$ & $0.02354$ \\
$74$ & L & $73$ & $2{,}701$ & $0$ & $0.02651$ & $75$ & Gmg & $75$ & $2{,}775$ & $30.45974$ & $0.02637$ \\
$76$ & Gmg & $75$ & $2{,}850$ & $31.82737$ & $0.02745$ & $77$ & Gmg & $77$ & $2{,}926$ & $30.75344$ & $0.02825$ \\
$78$ & L & $77$ & $3{,}003$ & $0$ & $0.02811$ & $79$ & Gmg & $79$ & $3{,}081$ & $36.80511$ & $0.03050$ \\
$80$ & P & $79$ & $3{,}160$ & $2.26216$ & $0.03355$ & $81$ & A & $81$ & $3{,}240$ & $1.66391$ & $0.42276$ \\
$82$ & L & $81$ & $3{,}321$ & $0$ & $0.03440$ & $83$ & Gmg & $83$ & $3{,}403$ & $63.71307$ & $0.03662$ \\
$84$ & Gmg & $83$ & $3{,}486$ & $270.91089$ & $0.03759$ & $85$ & Gmg & $85$ & $3{,}570$ & $72.56609$ & $0.04090$ \\
$86$ & L & $85$ & $3{,}655$ & $0$ & $0.04138$ & $87$ & Gmg & $87$ & $3{,}741$ & $107.21964$ & $0.04492$ \\
$88$ & Gmg & $87$ & $3{,}828$ & $29.36341$ & $0.04650$ & $89$ & Gmg & $89$ & $3{,}916$ & $13.81405$ & $0.04693$ \\
$90$ & L & $89$ & $4{,}005$ & $0$ & $0.04573$ & $91$ & Gmg & $91$ & $4{,}095$ & $106.57981$ & $0.05440$ \\
$92$ & Gmg & $91$ & $4{,}186$ & $179.49719$ & $0.05337$ & $93$ & Gmg & $93$ & $4{,}278$ & $231.76871$ & $0.05561$ \\
$94$ & L & $93$ & $4{,}371$ & $0$ & $0.06615$ & $95$ & Gmg & $95$ & $4{,}465$ & $171.23421$ & $0.06089$ \\
$96$ & Gmg & $95$ & $4{,}560$ & $101.56321$ & $0.06414$ & $97$ & Gmg & $97$ & $4{,}656$ & $48.14166$ & $0.06838$ \\
$98$ & L & $97$ & $4{,}753$ & $0$ & $0.06119$ & $99$ & P & $99$ & $4{,}851$ & $3.89104$ & $0.07727$ \\
$100$ & A & $99$ & $4{,}950$ & $2.14200$ & $0.22710$ & $101$ & Gmg & $101$ & $5{,}050$ & $392.50587$ & $0.13961$ \\
$102$ & L & $101$ & $5{,}151$ & $0$ & $0.08199$ & $103$ & Gmg & $103$ & $5{,}253$ & $564.36513$ & $0.15780$ \\
$104$ & Gmg & $103$ & $5{,}356$ & $617.33964$ & $0.16018$ & $105$ & Gmg & $105$ & $5{,}460$ & $198.86688$ & $0.15461$ \\
$106$ & L & $105$ & $5{,}565$ & $0$ & $0.10038$ & $107$ & Gmg & $107$ & $5{,}671$ & $96.85938$ & $0.16790$ \\
$108$ & Gmg & $107$ & $5{,}778$ & $162.31386$ & $0.15791$ & $109$ & Gmg & $109$ & $5{,}886$ & $89.91303$ & $0.17962$ \\
$110$ & L & $109$ & $5{,}995$ & $0$ & $0.10437$ & $111$ & Gmg & $111$ & $6{,}105$ & $702.10383$ & $0.21229$ \\
$112$ & Gmg & $111$ & $6{,}216$ & $1{,}642.77767$ & $0.22564$ & $113$ & Gmg & $113$ & $6{,}328$ & $521.93720$ & $0.21527$ \\
$114$ & L & $113$ & $6{,}441$ & $0$ & $0.12483$ & $115$ & Gmg & $115$ & $6{,}555$ & $350.64985$ & $0.23880$ \\
$116$ & Gmg & $115$ & $6{,}670$ & $2{,}902.77547$ & $0.23939$ & $117$ & Gmg & $117$ & $6{,}786$ & $304.47080$ & $0.24672$ \\
$118$ & L & $117$ & $6{,}903$ & $0$ & $0.14847$ & $119$ & Gmg & $119$ & $7{,}021$ & $104.86010$ & $0.24462$ \\
$120$ & P & $119$ & $7{,}140$ & $3.03291$ & $0.27342$ & $121$ & A & $121$ & $7{,}260$ & $2.79227$ & $0.83946$ \\
$122$ & L & $121$ & $7{,}381$ & $0$ & $0.16289$ & $123$ & Gmg & $123$ & $7{,}503$ & $2{,}387.17665$ & $0.33385$ \\
$124$ & Gmg & $123$ & $7{,}626$ & $6{,}717.33498$ & $0.31586$ & $125$ & Gmg & $125$ & $7{,}750$ & $1{,}848.82540$ & $0.31987$ \\
$126$ & L & $125$ & $7{,}875$ & $0$ & $0.18785$ & $127$ & Gmg & $127$ & $8{,}001$ & $2{,}503.02294$ & $0.38856$ \\
$128$ & Gmg & $127$ & $8{,}128$ & $6{,}604.14848$ & $0.34950$ & $129$ & Gmg & $129$ & $8{,}256$ & $2{,}685.70456$ & $0.39875$ \\
$130$ & L & $129$ & $8{,}385$ & $0$ & $0.20150$ & $131$ & Gmg & $131$ & $8{,}515$ & $699.12216$ & $0.44905$ \\
$132$ & P & $131$ & $8{,}646$ & $43.45062$ & $0.41763$ & $133$ & Gmg & $133$ & $8{,}778$ & $3{,}086.83193$ & $0.46623$ \\
$134$ & L & $133$ & $8{,}911$ & $0$ & $0.22038$ & $135$ & Gmg & $135$ & $9{,}045$ & $7{,}623.78213$ & $0.47938$ \\
$136$ & Gmg & $135$ & $9{,}180$ & $6{,}845.27138$ & $0.44366$ & $137$ & Gmg & $137$ & $9{,}316$ & $5{,}829.07948$ & $0.46654$ \\
$138$ & L & $137$ & $9{,}453$ & $0$ & $0.28486$ & $139$ & Gmg & $139$ & $9{,}591$ & $4{,}723.43336$ & $0.52933$ \\
$140$ & Gmg & $139$ & $9{,}730$ & $4{,}206.50204$ & $0.50009$ & $141$ & Gmg & $141$ & $9{,}870$ & $7{,}892.86346$ & $0.54256$ \\
$142$ & L & $141$ & $10{,}011$ & $0$ & $0.25984$ & $143$ & P & $143$ & $10{,}153$ & $4.42164$ & $0.64923$ \\
$144$ & A & $143$ & $10{,}296$ & $9.45186$ & $1.08597$ & $145$ & Gmg & $145$ & $10{,}440$ & $6{,}005.63791$ & $0.60302$ \\
$146$ & L & $145$ & $10{,}585$ & $0$ & $0.34719$ & $147$ & Gmg & $147$ & $10{,}731$ & $2{,}798.57715$ & $0.58261$ \\
$148$ & Gmg & $147$ & $10{,}878$ & $17{,}367.30739$ & $0.67009$ & $149$ & Gmg & $149$ & $11{,}026$ & $12{,}013.46039$ & $0.67278$ \\
$150$ & L & $149$ & $11{,}175$ & $0$ & $0.33064$ & $151$ & Gmg & $151$ & $11{,}325$ & $4{,}361.83573$ & $0.73832$ \\
$152$ & Gmg & $151$ & $11{,}476$ & $21{,}131.84073$ & $0.79018$ & $153$ & Gmg & $153$ & $11{,}628$ & $4{,}027.27203$ & $0.76019$ \\
$154$ & L & $153$ & $11{,}781$ & $0$ & $0.37400$ & $155$ & Gmg & $155$ & $11{,}935$ & $1{,}071.67261$ & $0.81107$ \\
$156$ & P & $155$ & $12{,}090$ & $904.75948$ & $0.74407$ & $157$ & Gmg & $157$ & $12{,}246$ & $11{,}129.25423$ & $0.90723$ \\
$158$ & L & $157$ & $12{,}403$ & $0$ & $0.51467$ & $159$ & Gmg & $159$ & $12{,}561$ & $7{,}724.73954$ & $0.84441$ \\
$160$ & Gmg & $159$ & $12{,}720$ & $28{,}997.38050$ & $0.88579$ & $161$ & Gmg & $161$ & $12{,}880$ & $5{,}061.61215$ & $1.01331$ \\
$162$ & L & $161$ & $13{,}041$ & $0$ & $0.45274$ & $163$ & Gmg & $163$ & $13{,}203$ & $4{,}672.80779$ & $1.04698$ \\
$164$ & Gmg & $163$ & $13{,}366$ & $12{,}347.81489$ & $1.04159$ & $165$ & Gmg & $165$ & $13{,}530$ & $8{,}484.89941$ & $1.15169$ \\
$166$ & L & $165$ & $13{,}695$ & $0$ & $0.54714$ & $167$ & Gmg & $167$ & $13{,}861$ & $11{,}930.67325$ & $1.22587$ \\
$168$ & P & $167$ & $14{,}028$ & $36.01993$ & $1.21453$ & $169$ & A & $169$ & $14{,}196$ & $8.63554$ & $1.98278$ \\
$170$ & L & $169$ & $14{,}365$ & $0$ & $0.62464$ & $171$ & Gmg & $171$ & $14{,}535$ & $13{,}465.87786$ & $1.34790$ \\
$172$ & Gmg & $171$ & $14{,}706$ & $5{,}827.44147$ & $1.59431$ & $173$ & Gmg & $173$ & $14{,}878$ & $10{,}316.03906$ & $1.33608$ \\
$174$ & L & $173$ & $15{,}051$ & $0$ & $0.68977$ & $175$ & Gmg & $175$ & $15{,}225$ & $13{,}811.82872$ & $1.18988$ \\
$176$ & Gmg & $175$ & $15{,}400$ & $14{,}850.36257$ & $1.50147$ & $177$ & Gmg & $177$ & $15{,}576$ & $4{,}634.72723$ & $1.59953$ \\
$178$ & L & $177$ & $15{,}753$ & $0$ & $0.70396$ & $179$ & Gmg & $179$ & $15{,}931$ & $12{,}394.18658$ & $1.55712$ \\
$180$ & Gmg & $179$ & $16{,}110$ & $6{,}992.65714$ & $1.58332$ & $181$ & Gmg & $181$ & $16{,}290$ & $2{,}956.95314$ & $1.67050$ \\
$182$ & L & $181$ & $16{,}471$ & $0$ & $0.90521$ & $183$ & Gmg & $183$ & $16{,}653$ & $29{,}466.16099$ & $1.73451$ \\
$184$ & Gmg & $183$ & $16{,}836$ & $19{,}718.35537$ & $1.81714$ & $185$ & Gmg & $185$ & $17{,}020$ & $16{,}464.02646$ & $1.75780$ \\
$186$ & L & $185$ & $17{,}205$ & $0$ & $1.06574$ & $187$ & Gmg & $187$ & $17{,}391$ & $6{,}474.50105$ & $1.98361$ \\
$188$ & Gmg & $187$ & $17{,}578$ & $17{,}908.44225$ & $1.98051$ & $189$ & Gmg & $189$ & $17{,}766$ & $17{,}993.54735$ & $2.21475$ \\
$190$ & L & $189$ & $17{,}955$ & $0$ & $1.21145$ & $191$ & Gmg & $191$ & $18{,}145$ & $36{,}645.80370$ & $2.42831$ \\
$192$ & Gmg & $191$ & $18{,}336$ & $19{,}524.32388$ & $2.25642$ & $193$ & Gmg & $193$ & $18{,}528$ & $2{,}436.72167$ & $2.23150$ \\
$194$ & L & $193$ & $18{,}721$ & $0$ & $1.21055$ & $195$ & P & $195$ & $18{,}915$ & $20.82835$ & $2.58654$ \\
$196$ & A & $195$ & $19{,}110$ & $4.64333$ & $3.01674$ & $197$ & Gmg & $197$ & $19{,}306$ & $315{,}972.17677$ & $3.59453$ \\
$198$ & L & $197$ & $19{,}503$ & $0$ & $1.28043$ & $199$ & Gmg & $199$ & $19{,}701$ & $29{,}157.74894$ & $2.77768$ \\
$200$ & Gmg & $199$ & $19{,}900$ & $90{,}495.73264$ & $1.88516$ & $201$ & Gmg & $201$ & $20{,}100$ & $175{,}275.15455$ & $2.30823$ \\
$202$ & L & $201$ & $20{,}301$ & $0$ & $1.41915$ & $203$ & Gmg & $203$ & $20{,}503$ & $57{,}207.26134$ & $1.85203$ \\
$204$ & Gmg & $203$ & $20{,}706$ & $114{,}385.07923$ & $4.27021$ & $205$ & Gmg & $205$ & $20{,}910$ & $73{,}983.76988$ & $4.48535$ \\
$206$ & L & $205$ & $21{,}115$ & $0$ & $1.62179$ & $207$ & Gmg & $207$ & $21{,}321$ & $77{,}422.04427$ & $3.33211$ \\
$208$ & Gmg & $207$ & $21{,}528$ & $79{,}593.92415$ & $4.29372$ & $209$ & Gmg & $209$ & $21{,}736$ & $9{,}203.88115$ & $3.11224$ \\
$210$ & L & $209$ & $21{,}945$ & $0$ & $1.83951$ & $211$ & Gmg & $211$ & $22{,}155$ & $99{,}840.40137$ & $4.51981$ \\
$212$ & Gmg & $211$ & $22{,}366$ & $117{,}210.05705$ & $3.26717$ & $213$ & Gmg & $213$ & $22{,}578$ & $65{,}807.32372$ & $3.97001$ \\
$214$ & L & $213$ & $22{,}791$ & $0$ & $1.99341$ & $215$ & Gmg & $215$ & $23{,}005$ & $53{,}534.75094$ & $2.92124$ \\
$216$ & Gmg & $215$ & $23{,}220$ & $132{,}836.32883$ & $5.70287$ & $217$ & Gmg & $217$ & $23{,}436$ & $172{,}825.28076$ & $4.84917$ \\
$218$ & L & $217$ & $23{,}653$ & $0$ & $2.05029$ & $219$ & Gmg & $219$ & $23{,}871$ & $49{,}951.16657$ & $2.96091$ \\
$220$ & Gmg & $219$ & $24{,}090$ & $168{,}645.38140$ & $2.82742$ & $221$ & Gmg & $221$ & $24{,}310$ & $189{,}407.89700$ & $3.36342$ \\
$222$ & L & $221$ & $24{,}531$ & $0$ & $2.30139$ & $223$ & Gmg & $223$ & $24{,}753$ & $58{,}344.26543$ & $9.49514$ \\
$224$ & P & $223$ & $24{,}976$ & $124.00168$ & $7.80894$ & $225$ & A & $225$ & $25{,}200$ & $95.16915$ & $6.11686$ \\
$226$ & L & $225$ & $25{,}425$ & $0$ & $2.46676$ & $227$ & Gmg & $227$ & $25{,}651$ & $276{,}323.98799$ & $4.87116$ \\
$228$ & Gmg & $227$ & $25{,}878$ & $183{,}727.48550$ & $3.66284$ & $229$ & Gmg & $229$ & $26{,}106$ & $80{,}441.63424$ & $4.90306$ \\
$230$ & L & $229$ & $26{,}335$ & $0$ & $2.43773$ & $231$ & Gmg & $231$ & $26{,}565$ & $143{,}986.63977$ & $5.08607$ \\
$232$ & Gmg & $231$ & $26{,}796$ & $189{,}898.36999$ & $6.91878$ & $233$ & Gmg & $233$ & $27{,}028$ & $121{,}375.52512$ & $7.69863$ \\
$234$ & L & $233$ & $27{,}261$ & $0$ & $2.69701$ & $235$ & Gmg & $235$ & $27{,}495$ & $136{,}858.38425$ & $6.43056$ \\
$236$\footnote{The depth-$235$ circuit was found by one of four independently seeded Gmg searches after the original search was stopped at $180$ complete independent layers.  Total search includes the original search and all four new threads; the five histories are shown in \fig{n236searches}.} & Gmg & $235$ & $27{,}730$ & $1{,}572{,}884.26193$ & $10.06714$ & $237$ & Gmg & $237$ & $27{,}966$ & $83{,}040.40247$ & $4.99658$ \\
$238$ & L & $237$ & $28{,}203$ & $0$ & $3.23544$ & $239$ & Gmg & $239$ & $28{,}441$ & $27{,}288.28900$ & $9.99705$ \\
$240$ & P & $239$ & $28{,}680$ & $205.76085$ & $11.12358$ & $241$ & Gmg & $241$ & $28{,}920$ & $438{,}803.87809$ & $7.71265$ \\
$242$ & L & $241$ & $29{,}161$ & $0$ & $3.57251$ & $243$ & Gmg & $243$ & $29{,}403$ & $354{,}758.21033$ & $11.89405$ \\
$244$ & Gmg & $243$ & $29{,}646$ & $342{,}667.14581$ & $5.52728$ & $245$ & Gmg & $245$ & $29{,}890$ & $94{,}323.63384$ & $5.87693$ \\
$246$ & L & $245$ & $30{,}135$ & $0$ & $3.76704$ & $247$ & Gmg & $247$ & $30{,}381$ & $148{,}838.66784$ & $9.65502$ \\
$248$ & Gmg & $247$ & $30{,}628$ & $284{,}180.30950$ & $10.32278$ & $249$ & Gmg & $249$ & $30{,}876$ & $265{,}757.16082$ & $8.35786$ \\
$250$ & L & $249$ & $31{,}125$ & $0$ & $3.54186$ & $251$ & Gmg & $251$ & $31{,}375$ & $183{,}419.07448$ & $11.05665$ \\
$252$\footnote{The depth-$251$ circuit was found by one of four independently seeded Gmg reruns.  Total search includes the original search, which was stopped at $197$ complete independent layers, and all four reruns; see \fig{n252searches}.} & Gmg & $251$ & $31{,}626$ & $1{,}678{,}680.78375$ & $15.80003$ & $253$ & Gmg & $253$ & $31{,}878$ & $128{,}898.70847$ & $8.52935$ \\
$254$ & L & $253$ & $32{,}131$ & $0$ & $5.59725$ & $255$ & P & $255$ & $32{,}385$ & $1{,}742.92704$ & $12.97873$ \\
$256$ & A & $255$ & $32{,}640$ & $123.47433$ & $12.05221$ & $258$ & L & $257$ & $33{,}153$ & $0$ & $4.22479$ \\
$262$ & L & $261$ & $34{,}191$ & $0$ & $5.89625$ & $266$ & L & $265$ & $35{,}245$ & $0$ & $5.69879$ \\
$270$ & L & $269$ & $36{,}315$ & $0$ & $6.50871$ & $272$ & P & $271$ & $36{,}856$ & $1{,}521.04525$ & $27.19428$ \\
$274$ & L & $273$ & $37{,}401$ & $0$ & $7.08797$ & $278$ & L & $277$ & $38{,}503$ & $0$ & $7.89822$ \\
$282$ & L & $281$ & $39{,}621$ & $0$ & $8.24725$ & $286$ & L & $285$ & $40{,}755$ & $0$ & $8.87482$ \\
$288$ & P & $287$ & $41{,}328$ & $371.97941$ & $29.55890$ & $289$ & A & $289$ & $41{,}616$ & $45.12759$ & $24.35325$ \\
$290$ & L & $289$ & $41{,}905$ & $0$ & $10.38025$ & $294$ & L & $293$ & $43{,}071$ & $0$ & $10.44373$ \\
$298$ & L & $297$ & $44{,}253$ & $0$ & $11.54227$ & $302$ & L & $301$ & $45{,}451$ & $0$ & $10.14563$ \\
$306$ & L & $305$ & $46{,}665$ & $0$ & $11.85909$ & $310$ & L & $309$ & $47{,}895$ & $0$ & $13.98466$ \\
$314$ & L & $313$ & $49{,}141$ & $0$ & $13.93340$ & $318$ & L & $317$ & $50{,}403$ & $0$ & $15.43658$ \\
$322$ & L & $321$ & $51{,}681$ & $0$ & $17.51178$ & $323$ & P & $323$ & $52{,}003$ & $3{,}725.78125$ & $52.05652$ \\
$324$ & A & $323$ & $52{,}326$ & $257.20074$ & $38.39955$ & $326$ & L & $325$ & $52{,}975$ & $0$ & $18.29986$ \\
$330$ & L & $329$ & $54{,}285$ & $0$ & $19.29247$ & $334$ & L & $333$ & $55{,}611$ & $0$ & $19.48430$ \\
$338$ & L & $337$ & $56{,}953$ & $0$ & $21.34158$ & $342$ & L & $341$ & $58{,}311$ & $0$ & $23.69139$ \\
$346$ & L & $345$ & $59{,}685$ & $0$ & $24.38477$ & $350$ & L & $349$ & $61{,}075$ & $0$ & $27.78790$ \\
$354$ & L & $353$ & $62{,}481$ & $0$ & $32.91005$ & $358$ & L & $357$ & $63{,}903$ & $0$ & $31.28075$ \\
$360$ & P & $359$ & $64{,}620$ & $1{,}582.22595$ & $99.43643$ & $361$ & A & $361$ & $64{,}980$ & $679.83494$ & $107.97132$ \\
$362$ & L & $361$ & $65{,}341$ & $0$ & $37.58978$ & $366$ & L & $365$ & $66{,}795$ & $0$ & $38.06586$ \\
$370$ & L & $369$ & $68{,}265$ & $0$ & $35.32630$ & $374$ & L & $373$ & $69{,}751$ & $0$ & $42.06352$ \\
$378$ & L & $377$ & $71{,}253$ & $0$ & $40.28716$ & $380$ & P & $379$ & $72{,}010$ & $40{,}315.54814$ & $97.50165$ \\
$382$ & L & $381$ & $72{,}771$ & $0$ & $45.63655$ & $386$ & L & $385$ & $74{,}305$ & $0$ & $46.59988$ \\
$390$ & L & $389$ & $75{,}855$ & $0$ & $52.28142$ & $394$ & L & $393$ & $77{,}421$ & $0$ & $59.72070$ \\
$398$ & L & $397$ & $79{,}003$ & $0$ & $56.16899$ & $399$ & P & $399$ & $79{,}401$ & $4{,}474.84888$ & $136.55566$ \\
$400$ & A & $399$ & $79{,}800$ & $681.32963$ & $175.21646$ & $402$ & L & $401$ & $80{,}601$ & $0$ & $55.28932$ \\
$406$ & L & $405$ & $82{,}215$ & $0$ & $64.00741$ & $410$ & L & $409$ & $83{,}845$ & $0$ & $55.55485$ \\
$414$ & L & $413$ & $85{,}491$ & $0$ & $71.98400$ & $418$ & L & $417$ & $87{,}153$ & $0$ & $60.76727$ \\
$420$ & P & $419$ & $87{,}990$ & $712.26806$ & $268.57933$ & $422$ & L & $421$ & $88{,}831$ & $0$ & $76.10127$ \\
$426$ & L & $425$ & $90{,}525$ & $0$ & $71.78816$ & $430$ & L & $429$ & $92{,}235$ & $0$ & $78.64478$ \\
$434$ & L & $433$ & $93{,}961$ & $0$ & $83.92123$ & $438$ & L & $437$ & $95{,}703$ & $0$ & $86.00190$ \\
$440$ & P & $439$ & $96{,}580$ & $6{,}662.06095$ & $217.30798$ & $441$ & A & $441$ & $97{,}020$ & $1{,}338.75039$ & $211.23053$ \\
$442$ & L & $441$ & $97{,}461$ & $0$ & $89.06298$ & $446$ & L & $445$ & $99{,}235$ & $0$ & $90.70318$ \\
$450$ & L & $449$ & $101{,}025$ & $0$ & $95.15723$ & $454$ & L & $453$ & $102{,}831$ & $0$ & $109.13572$ \\
$458$ & L & $457$ & $104{,}653$ & $0$ & $109.22388$ & $462$ & L & $461$ & $106{,}491$ & $0$ & $119.37494$ \\
$466$ & L & $465$ & $108{,}345$ & $0$ & $115.09336$ & $470$ & L & $469$ & $110{,}215$ & $0$ & $119.39378$ \\
$474$ & L & $473$ & $112{,}101$ & $0$ & $136.13666$ & $478$ & L & $477$ & $114{,}003$ & $0$ & $147.62243$ \\
$482$ & L & $481$ & $115{,}921$ & $0$ & $146.40654$ & $483$ & P & $483$ & $116{,}403$ & $17{,}107.39612$ & $431.57289$ \\
$484$ & A & $483$ & $116{,}886$ & $3{,}071.93487$ & $385.14724$ & $486$ & L & $485$ & $117{,}855$ & $0$ & $150.76992$ \\
$490$ & L & $489$ & $119{,}805$ & $0$ & $166.59270$ & $494$ & L & $493$ & $121{,}771$ & $0$ & $145.95390$ \\
$498$ & L & $497$ & $123{,}753$ & $0$ & $176.80330$ & $502$ & L & $501$ & $125{,}751$ & $0$ & $191.42964$ \\
$506$ & L & $505$ & $127{,}765$ & $0$ & $195.66624$ & $510$ & L & $509$ & $129{,}795$ & $0$ & $93.32939$ \\
$514$ & L & $513$ & $131{,}841$ & $0$ & $95.73907$ & $518$ & L & $517$ & $133{,}903$ & $0$ & $97.57528$ \\
$522$ & L & $521$ & $135{,}981$ & $0$ & $129.31070$ & $526$ & L & $525$ & $138{,}075$ & $0$ & $111.21035$ \\
$528$ & P & $527$ & $139{,}128$ & $30{,}973.22326$ & $843.55639$ & $529$ & A & $529$ & $139{,}656$ & $10{,}015.07655$ & $708.08168$ \\
$530$ & L & $529$ & $140{,}185$ & $0$ & $120.00808$ & $534$ & L & $533$ & $142{,}311$ & $0$ & $127.54124$ \\
$538$ & L & $537$ & $144{,}453$ & $0$ & $133.47928$ & $542$ & L & $541$ & $146{,}611$ & $0$ & $148.72254$ \\
$546$ & L & $545$ & $148{,}785$ & $0$ & $145.49512$ & $550$ & L & $549$ & $150{,}975$ & $0$ & $169.34354$ \\
$552$ & P & $551$ & $152{,}076$ & $468{,}800.60380$ & $561.38637$ & $554$ & L & $553$ & $153{,}181$ & $0$ & $185.69865$ \\
$558$ & L & $557$ & $155{,}403$ & $0$ & $225.77579$ & $562$ & L & $561$ & $157{,}641$ & $0$ & $247.69023$ \\
$566$ & L & $565$ & $159{,}895$ & $0$ & $218.98687$ & $570$ & L & $569$ & $162{,}165$ & $0$ & $254.21911$ \\
$574$ & L & $573$ & $164{,}451$ & $0$ & $282.78337$ & $575$ & P & $575$ & $165{,}025$ & $16{,}803.64963$ & $1{,}004.20631$ \\
$576$ & A & $575$ & $165{,}600$ & $2{,}923.25988$ & $1{,}338.91268$ & $578$ & L & $577$ & $166{,}753$ & $0$ & $297.09850$ \\
$582$ & L & $581$ & $169{,}071$ & $0$ & $325.16798$ & $586$ & L & $585$ & $171{,}405$ & $0$ & $335.50599$ \\
$590$ & L & $589$ & $173{,}755$ & $0$ & $281.56106$ & $594$ & L & $593$ & $176{,}121$ & $0$ & $344.50434$ \\
$598$ & L & $597$ & $178{,}503$ & $0$ & $365.62265$ & $602$ & L & $601$ & $180{,}901$ & $0$ & $344.08141$ \\
$606$ & L & $605$ & $183{,}315$ & $0$ & $396.66236$ & $610$ & L & $609$ & $185{,}745$ & $0$ & $420.07916$ \\
$614$ & L & $613$ & $188{,}191$ & $0$ & $438.82567$ & $618$ & L & $617$ & $190{,}653$ & $0$ & $475.95695$ \\
$622$ & L & $621$ & $193{,}131$ & $0$ & $517.77495$ & $625$ & A & $625$ & $195{,}000$ & $25{,}903.36288$ & $1{,}664.11884$ \\
$626$ & L & $625$ & $195{,}625$ & $0$ & $585.18200$ & $630$ & L & $629$ & $198{,}135$ & $0$ & $509.39107$ \\
$634$ & L & $633$ & $200{,}661$ & $0$ & $537.28545$ & $638$ & L & $637$ & $203{,}203$ & $0$ & $577.12304$ \\
$642$ & L & $641$ & $205{,}761$ & $0$ & $601.96093$ & $646$ & L & $645$ & $208{,}335$ & $0$ & $565.07189$ \\
$650$ & L & $649$ & $210{,}925$ & $0$ & $295.54689$ & $654$ & L & $653$ & $213{,}531$ & $0$ & $363.27194$ \\
$658$ & L & $657$ & $216{,}153$ & $0$ & $389.24336$ & $662$ & L & $661$ & $218{,}791$ & $0$ & $367.45362$ \\
$666$ & L & $665$ & $221{,}445$ & $0$ & $716.28932$ & $670$ & L & $669$ & $224{,}115$ & $0$ & $740.25678$ \\
$674$ & L & $673$ & $226{,}801$ & $0$ & $885.63651$ & $675$ & P & $675$ & $227{,}475$ & $59{,}911.00657$ & $3{,}368.61720$ \\
$676$ & A & $675$ & $228{,}150$ & $25{,}280.35053$ & $2{,}413.83203$ & $678$ & L & $677$ & $229{,}503$ & $0$ & $1{,}023.09529$ \\
$682$ & L & $681$ & $232{,}221$ & $0$ & $842.31679$ & $686$ & L & $685$ & $234{,}955$ & $0$ & $1{,}101.27608$ \\
$690$ & L & $689$ & $237{,}705$ & $0$ & $1{,}034.37345$ & $694$ & L & $693$ & $240{,}471$ & $0$ & $1{,}072.01650$ \\
$698$ & L & $697$ & $243{,}253$ & $0$ & $978.19860$ & $702$ & L & $701$ & $246{,}051$ & $0$ & $616.28452$ \\
$706$ & L & $705$ & $248{,}865$ & $0$ & $1{,}092.30869$ & $710$ & L & $709$ & $251{,}695$ & $0$ & $1{,}210.80668$ \\
$714$ & L & $713$ & $254{,}541$ & $0$ & $1{,}030.08741$ & $718$ & L & $717$ & $257{,}403$ & $0$ & $968.65959$ \\
$722$ & L & $721$ & $260{,}281$ & $0$ & $1{,}179.04384$ & $726$ & L & $725$ & $263{,}175$ & $0$ & $1{,}173.60612$ \\
$728$ & P & $727$ & $264{,}628$ & $206{,}822.30297$ & $3{,}930.61300$ & $729$ & A & $729$ & $265{,}356$ & $94{,}993.18209$ & $3{,}779.45965$ \\
$730$ & L & $729$ & $266{,}085$ & $0$ & $1{,}000.40460$ & $734$ & L & $733$ & $269{,}011$ & $0$ & $940.00523$ \\
$738$ & L & $737$ & $271{,}953$ & $0$ & $639.16926$ & $742$ & L & $741$ & $274{,}911$ & $0$ & $1{,}082.24413$ \\
$746$ & L & $745$ & $277{,}885$ & $0$ & $1{,}281.51952$ & $750$ & L & $749$ & $280{,}875$ & $0$ & $1{,}321.65807$ \\
$754$ & L & $753$ & $283{,}881$ & $0$ & $709.40361$ & $756$ & P & $755$ & $285{,}390$ & $225{,}805.74245$ & $7{,}268.17288$ \\
$758$ & L & $757$ & $286{,}903$ & $0$ & $993.95252$ & $762$ & L & $761$ & $289{,}941$ & $0$ & $738.05799$ \\
$766$ & L & $765$ & $292{,}995$ & $0$ & $1{,}431.95478$ & $770$ & L & $769$ & $296{,}065$ & $0$ & $1{,}466.07925$ \\
$774$ & L & $773$ & $299{,}151$ & $0$ & $789.53991$ & $778$ & L & $777$ & $302{,}253$ & $0$ & $810.15806$ \\
$782$ & L & $781$ & $305{,}371$ & $0$ & $892.37272$ & $783$ & P & $783$ & $306{,}153$ & $438{,}036.62126$ & $6{,}691.90781$ \\
$784$ & A & $783$ & $306{,}936$ & $20{,}360.76683$ & $5{,}977.33692$ & $786$ & L & $785$ & $308{,}505$ & $0$ & $1{,}357.81227$ \\
$790$ & L & $789$ & $311{,}655$ & $0$ & $2{,}161.01737$ & $794$ & L & $793$ & $314{,}821$ & $0$ & $2{,}196.71059$ \\
$798$ & L & $797$ & $318{,}003$ & $0$ & $1{,}930.30033$ & $802$ & L & $801$ & $321{,}201$ & $0$ & $911.49596$ \\
$806$ & L & $805$ & $324{,}415$ & $0$ & $1{,}271.64568$ & $810$ & L & $809$ & $327{,}645$ & $0$ & $1{,}442.30302$ \\
$812$ & P & $811$ & $329{,}266$ & $399{,}054.03207$ & $8{,}872.83003$ & $814$ & L & $813$ & $330{,}891$ & $0$ & $1{,}260.31304$ \\
$818$ & L & $817$ & $334{,}153$ & $0$ & $1{,}698.74901$ & $822$ & L & $821$ & $337{,}431$ & $0$ & $1{,}385.07312$ \\
$826$ & L & $825$ & $340{,}725$ & $0$ & $1{,}549.58674$ & $830$ & L & $829$ & $344{,}035$ & $0$ & $1{,}724.32998$ \\
$834$ & L & $833$ & $347{,}361$ & $0$ & $1{,}690.16606$ & $838$ & L & $837$ & $350{,}703$ & $0$ & $1{,}456.57692$ \\
$840$ & P & $839$ & $352{,}380$ & $204{,}870.01771$ & $11{,}197.40698$ & $841$ & A & $841$ & $353{,}220$ & $14{,}963.89367$ & $7{,}781.78329$ \\
$842$ & L & $841$ & $354{,}061$ & $0$ & $1{,}358.08728$ & $846$ & L & $845$ & $357{,}435$ & $0$ & $1{,}810.78340$ \\
$850$ & L & $849$ & $360{,}825$ & $0$ & $1{,}296.43126$ & $854$ & L & $853$ & $364{,}231$ & $0$ & $1{,}493.48309$ \\
$858$ & L & $857$ & $367{,}653$ & $0$ & $1{,}433.35759$ & $862$ & L & $861$ & $371{,}091$ & $0$ & $2{,}078.99568$ \\
$866$ & L & $865$ & $374{,}545$ & $0$ & $1{,}454.61985$ & $870$ & L & $869$ & $378{,}015$ & $0$ & $1{,}889.66705$ \\
$874$ & L & $873$ & $381{,}501$ & $0$ & $1{,}439.20568$ & $878$ & L & $877$ & $385{,}003$ & $0$ & $2{,}330.44248$ \\
$882$ & L & $881$ & $388{,}521$ & $0$ & $1{,}595.39843$ & $886$ & L & $885$ & $392{,}055$ & $0$ & $2{,}437.16942$ \\
$890$ & L & $889$ & $395{,}605$ & $0$ & $1{,}612.10907$ & $894$ & L & $893$ & $399{,}171$ & $0$ & $2{,}550.01701$ \\
$898$ & L & $897$ & $402{,}753$ & $0$ & $2{,}731.96542$ & $900$ & A & $899$ & $404{,}550$ & $70{,}097.08027$ & $16{,}868.25791$ \\
$902$ & L & $901$ & $406{,}351$ & $0$ & $3{,}037.66058$ & $906$ & L & $905$ & $409{,}965$ & $0$ & $2{,}343.59945$ \\
$910$ & L & $909$ & $413{,}595$ & $0$ & $2{,}862.47465$ & $914$ & L & $913$ & $417{,}241$ & $0$ & $2{,}553.69583$ \\
$918$ & L & $917$ & $420{,}903$ & $0$ & $2{,}311.66584$ & $922$ & L & $921$ & $424{,}581$ & $0$ & $1{,}638.87264$ \\
$926$ & L & $925$ & $428{,}275$ & $0$ & $2{,}472.99139$ & $930$ & L & $929$ & $431{,}985$ & $0$ & $2{,}467.82280$ \\
$934$ & L & $933$ & $435{,}711$ & $0$ & $2{,}186.89441$ & $938$ & L & $937$ & $439{,}453$ & $0$ & $3{,}471.64637$ \\
$942$ & L & $941$ & $443{,}211$ & $0$ & $2{,}504.29328$ & $946$ & L & $945$ & $446{,}985$ & $0$ & $2{,}376.53662$ \\
$950$ & L & $949$ & $450{,}775$ & $0$ & $4{,}093.24908$ & $954$ & L & $953$ & $454{,}581$ & $0$ & $3{,}358.20086$ \\
$958$ & L & $957$ & $458{,}403$ & $0$ & $4{,}790.32414$ & $960$ & P & $959$ & $460{,}320$ & $197{,}414.01554$ & $25{,}810.84587$ \\
$961$ & A & $961$ & $461{,}280$ & $29{,}308.22829$ & $28{,}933.05388$ & $962$ & L & $961$ & $462{,}241$ & $0$ & $2{,}668.07471$ \\
$966$ & L & $965$ & $466{,}095$ & $0$ & $3{,}706.38495$ & $970$ & L & $969$ & $469{,}965$ & $0$ & $2{,}659.40492$ \\
$974$ & L & $973$ & $473{,}851$ & $0$ & $2{,}790.95532$ & $978$ & L & $977$ & $477{,}753$ & $0$ & $4{,}441.78180$ \\
$982$ & L & $981$ & $481{,}671$ & $0$ & $3{,}396.57008$ & $986$ & L & $985$ & $485{,}605$ & $0$ & $2{,}981.18064$ \\
$990$ & L & $989$ & $489{,}555$ & $0$ & $3{,}195.76309$ & $994$ & L & $993$ & $493{,}521$ & $0$ & $3{,}711.81897$ \\
$998$ & L & $997$ & $497{,}503$ & $0$ & $4{,}187.22864$ & $1{,}002$ & L & $1{,}001$ & $501{,}501$ & $0$ & $3{,}345.16420$ \\
$1{,}006$ & L & $1{,}005$ & $505{,}515$ & $0$ & $3{,}554.19046$ & $1{,}010$ & L & $1{,}009$ & $509{,}545$ & $0$ & $2{,}651.44396$ \\
$1{,}014$ & L & $1{,}013$ & $513{,}591$ & $0$ & $3{,}531.07295$ & $1{,}018$ & L & $1{,}017$ & $517{,}653$ & $0$ & $3{,}726.84154$ \\
$1{,}022$ & L & $1{,}021$ & $521{,}731$ & $0$ & $4{,}760.64194$ & $1{,}023$ & P & $1{,}023$ & $522{,}753$ & $72{,}492.08742$ & $33{,}631.32152$ \\
$1{,}024$ & A & $1{,}023$ & $523{,}776$ & $63{,}708.99658$ & $33{,}721.61545$ &  &  &  &  &  &  \\

\multicolumn{12}{c}{}\\
\end{longtable}
\endgroup
\clearpage

\section{Clifford unitary synthesis}\label{sec:unitaries}

State preparation in \eq{endpointabsorb} absorbs an arbitrary linear reversible endpoint because the input is $\ket{+}^{\otimes n}$.  Unitary synthesis must instead act correctly on every input.  We first give a useful endpoint of the cycle walk and then show how to cancel general linear endpoints in pairs of programmed $\{\cnotgate,\pgate\}$ circuits.

\subsection{A diagonal circuit with a fixed half-turn}\label{ssec:halfturn}

\begin{lemma}[Power-of-two cycle closure]\label{lem:halfturn}
For $n\,{=}\,2^k\,{\geq}\,4$, an arbitrary $\czgate$ circuit can be implemented on $C_n$ in depth $3n/2$, up to the fixed cyclic shift by $n/2$ vertices.  The same holds on the minimum-perimeter patches of \lem{patches}.
\end{lemma}

\begin{proof}
Let $n\,{=}\,2m$.  By \eq{cyclecellshift} in \app{cycles}, $T^2\,{\oplus}\,T\,{=}\,R^{-2}$.  Since $m$ is a power of two, raising this identity to the $m$th power over $\FF_2$ gives $T^{2m}\,{\oplus}\,T^m\,{=}\,I$.  Multiplying by $T^m\,{\oplus}\,I$ yields $T^{3m}\,{=}\,I$.  By \eq{cycleprefix}, the endpoint at depth $3m$ is therefore $R^{3m}\,{=}\,R^m$, the half-turn.  The first $n$ layers already have full span by \lem{cycles}.
\end{proof}

\lem{halfturn} gives a standalone diagonal operation with a known output order, rather than only a state preparation.  Four alternating diagonal blocks would give Clifford depth $6n$ at these sizes (\eq{fourdiagonal}), since the four half-turns can be moved to the end and cancel.  The construction below gives a stronger bound for every even size without waiting for a permutation endpoint.

\subsection{Cancellation of linear endpoints}\label{ssec:pairing}

Let $\calD_Z$ be the group of diagonal Clifford operators generated by $\pgate$ and $\czgate$, and put $\calD_X\,{=}\,\hgate^{\otimes n}\calD_Z\hgate^{\otimes n}$.  Conjugation by a linear reversible operation $U_A$ preserves both groups.  For $\calD_Z$ this follows by substituting $Ax$ in its phase polynomial; for $\calD_X$ it follows from the identity
\begin{equation}\label{eq:fouriertranspose}
\hgate^{\otimes n} U_{A^{\mathsf T}}\hgate^{\otimes n}\,{=}\,U_{A^{-1}}.
\end{equation}
For a fixed $\cnotgate$ circuit $C$, define $C^{\mathsf t}$ by reversing the execution order and interchanging every control and target.  If $C$ computes the binary wire matrix $A$, then $C^{\mathsf t}$ computes $A^{\mathsf T}$, a transpose of the binary wire matrix.

\begin{lemma}[Paired diagonal stages]\label{lem:pairing}
Suppose that $C$ and $C^{\mathsf t}$ both have full $\czgate$ span, and that $C$ has depth $d$.  Any $XZ$, with $Z\,{\in}\,\calD_Z$ and $X\,{\in}\,\calD_X$, can be implemented exactly in two-qubit depth at most $2d$.
\end{lemma}

\begin{proof}
Put $X'\,{=}\,U_A XU_{A^{-1}}\,{\in}\,\calD_X$ and $Z'\,{=}\,\hgate^{\otimes n} X'\hgate^{\otimes n}\,{\in}\,\calD_Z$.  By \propos{rank}, program $C$ for $U_A Z$ and $C^{\mathsf t}$ for $U_{A^{\mathsf T}}Z'$.  Execute the first circuit, $\hgate^{\otimes n}$, the second circuit, and $\hgate^{\otimes n}$, in that order.  Using \eq{fouriertranspose}, their product is
\begin{equation}\label{eq:paircancellation}
\hgate^{\otimes n}(U_{A^{\mathsf T}}Z')\hgate^{\otimes n}(U_A Z)
\,{=}\,U_{A^{-1}}X'U_A Z\,{=}\,XZ.
\end{equation}
The two residual linear operations cancel.  
\end{proof}

The assumption about $C^{\mathsf t}$ matters: full span of an arbitrary circuit does not imply full span after transposition.  For the even-cycle circuit, it follows from two simple symmetries.  Reversing the execution order preserves the span rank: each reversed gate has the original pair-coefficient vector, with the common invertible change of input variables $A^{-1}$.  Reversing all arrows of the clockwise cycle circuit is equivalent to the relabeling $i\,{\mapsto}\,1{-}i\pmod n$, which preserves each of its two undirected matchings.  Thus both operations preserve its rank.  The searches in  \sec{search} can be modified to require that both $C$ and $C^{\mathsf t}$ span all $\czgate$ dimensions.  This appears to produce valid solutions for many small numbers $n$ tried, offering depth $4n-4$ Clifford circuits for $n{=}4k$ and depth $4n$ ancilla-free Clifford circuits for odd $n$ when a valid solution exists by following the construction in \lem{unitary}.

\subsection{Four programmed circuits suffice}\label{ssec:fourblocks}

An arbitrary Clifford unitary admits a circuit implementation with four alternating diagonal stages, up to local Clifford gates \cite{nemirovsky2026global}.  In operator order we may write
\begin{equation}\label{eq:fourdiagonal}
U\,{=}\,L_{\rm out}X_2Z_2X_1Z_1L_{\rm in},\qquad
X_1,X_2\,{\in}\,\calD_X,\quad Z_1,Z_2\,{\in}\,\calD_Z.
\end{equation}
Here $L_{\rm in},L_{\rm out}$ are local Clifford.  The published decomposition alternates programmable multiqubit interactions diagonal in the $X$ and $Z$ bases; their single-qubit phases belong to the same diagonal groups.  Conjugating by $\hgate^{\otimes n}$ interchanges the bases if the opposite starting order is used.  

\begin{lemma}[Clifford unitary depth on cycles and the square lattice]\label{lem:unitary}
For every even $n\,{\geq}\,4$, an arbitrary Clifford unitary can be implemented as a circuit on $C_n$, or on any $n$-vertex graph containing a Hamiltonian cycle, in the two-qubit depth at most
\begin{equation}\label{eq:unitarydepth}
\begin{cases}
4n{-}4,&n\,{=}\,4k{+}2,\\
4n,&n\,{=}\,4k.
\end{cases}
\end{equation}
In particular, these bounds hold on connected, hole-free minimum-perimeter 2DSL patches.  The circuit requires no ancillary qubits or measurements, and implements qubit permutation included in the specification of the Clifford unitary.
\end{lemma}

\begin{proof}
Take $C$ to be the depth-$d_C(n)$ cycle prefix in \lem{cycles}.  The preceding symmetry argument shows that $C^{\mathsf t}$ also has full span.  Apply \lem{pairing} twice to \eq{fourdiagonal}.  By \eq{paircancellation}, each pair has the identity linear endpoint, giving total depth $4d_C(n)$ and the bounds in \eq{unitarydepth}.  All gates remain on the cycle.  Finally use \lem{patches} for the square-lattice statement.
\end{proof}

This is a fixed universal programmable Clifford circuit: its four chronological $\cnotgate$ stages are $C,C^{\mathsf t},C,C^{\mathsf t}$, independent of the target unitary.  Only the Phase programs and the local Clifford endpoints depend on $U$.  The corresponding $\cnotgate$ count is $2nd_C(n)$, or $2n(n{-}1)$ and $2n^2$ in the two cases.  These upper bounds are furthermore constructive.

For comparison, the known ancilla-free LNN Clifford depth bound is $7n{-}4$ \cite{maslov2023cnot}.  Goubault de Brugi\`ere and Martiel give a $4n{+}O(1)$ grid construction for arbitrary $\cnotgate$ circuits \cite{goubault2023shallower}.  Their linear synthesis conditions do not require a full programmable $\czgate$ span.  Our result implements arbitrary Clifford unitaries, of which there are roughly quadratically more than linear invertible transformations, with the same leading coefficient of four by pairing full-span cycle circuits, without requiring an extension of their $\cnotgate$ construction.

\subsection{Odd sizes with one additional qubit}\label{ssec:oddunitaries}

For odd $n\,{\geq}\,3$, apply \lem{unitary} to $U\otimes I$ on $n{+}1$ vertices of a minimum-perimeter patch.  This gives depth at most $4n$ when $n\,{=}\,4k{+}1$ and $4n{+}4$ when $n\,{=}\,4k{+}3$.  

Indeed, in each $X_iZ_i$ pair, we compute both Phase programs for all $n{+}1$ variables simultaneously, including all terms involving the ancillary variable and the required Phase gates.  The two $\cnotgate$ schedules become inverses when the second is conjugated by its surrounding $\hgate$ layers.  With the Phase programs chosen accordingly, their residual linear transformations cancel on all $n{+}1$ qubits, including the ancillary qubit.  The completed pair therefore implements exactly $(X_iZ_i)\otimes I$, returning the ancillary qubit unchanged for any initial state.  One {\it dirty} ancilla therefore suffices to cover all odd sizes without a separate odd-size unitary construction.

\section{Conclusion}\label{sec:conclusion}

In this work, we focused on developing depth-optimized circuit implementations of diagonal Clifford operations, Clifford states, and arbitrary Clifford unitaries with nearest-neighbor gates over 2D Square Lattices.  

On a $2{\times}k$ ladder ($n\,{=}\,2k$), complementary reversal $\cnotgate$ gate walks expose every within-rail and cross-rail $\czgate$ interaction in depth at most $\frac{3n}{2}{+}3$. The construction reverses the order of qubits on both rails.  Compared with the best-known LNN-compliant depth-$(2n{+}2)$ construction \cite{maslov2018shorter}, this reduces the leading depth coefficient by $25\%$.

For state synthesis, the $\czgate$ rank criterion gives the counting lower bound $d_0(n):=2\lceil\frac{n}{2}\rceil \,{-}\,1$, together with strong matching restrictions at equality.  These restrictions exclude LNN and every tree on at least four vertices, as well as some low-diameter graphs such as the Petersen graph.  Closing an LNN path into a cycle requires only one additional edge, yet changes this conclusion: a cycle attains $d_0(n)$ whenever $n{=}\,4k{+}2$, and requires exactly one additional layer for $n\,{=}\,4k$.  A Hamiltonian minimum-perimeter, hole-free 2DSL patch exists for every even $n$.  Consequently, optimal-depth $\czgate$-universal state preparation on such patches is constructive for every $n\,{=}\,4k{+}2$; for $n\,{=}\,4k$, one additional $\cnotgate$ layer suffices.  The resulting depths $n{-}1$ and $n$ improve the comparable LNN depth bound $2n{+}2$ \cite{maslov2018shorter} by about $50\%$.

The search algorithms address the remaining cases beyond $n\,{=}\,4k{+}2$.  A and P search short periodic square and rectangular circuits; Gmg grows productive matching layers with geometric pruning and soft guidance.  The captured snapshot reported in \tab{allresults} contains $\ResultCount$ certified sizes, including every $n$ up to 256, and additional sizes through $\LargestN$.  All reported depths are optimal in the $\czgate$-universal programming model, including small-case exclusions at $n\,{=}\,4,8$.  Thus, computational evidence supports the existence of a minimal-depth solution beyond the analytic $4k{+}2$ family without establishing the respective results formally.

For arbitrary Clifford unitaries, pairing full-span cycle circuits with their Hadamard conjugations cancels the linear reversible endpoints.  Combined with the known four-diagonal-stage decomposition, this gives depth $4n{-}4$ for $n\,{=}\,4k{+}2$ and $4n$ for $n\,{=}\,4k$.  These bounds improve the best-known LNN construction of depth $7n{-}4$ \cite{maslov2023cnot} to $4n{+}O(1)$, reducing the leading depth coefficient by about $43\%$.  For even sizes, this improvement requires only the single edge that closes the LNN path into a cycle.  It is striking that one additional edge both overcomes the counting-depth obstruction for the state-programming families above and enables this reduction in general Clifford unitary depth.  One additional dirty qubit, returned unchanged, extends the $4n{+}O(1)$ unitary bound to all odd sizes.

Remaining questions include a general optimal-depth construction for the strict state-programming problem at odd sizes and at $n\,{=}\,4k$ and improved lower bounds for arbitrary Clifford unitary depth.  Optimizing individual states, including their local-complementation freedom, is a separate opportunity: initial Gmg experiments gave small consistent depth reductions, which we did not investigate in detail.

\section*{Acknowledgments}

I thank Pulkit Sinha from the University of Waterloo for helpful discussions. 

Codex GPT-6.1 Sol was a co-author of this work.  The collaboration included the development and checking of mathematical arguments, the design and implementation of search algorithms, computational experiments, independent verification, and the preparation of the manuscript.

\appendix
\section{Proof of the ladder span lemma}\label{app:ladder}

We prove \lem{AB} by changing variables on the two rails and canceling rectangles of $\czgate$ gates.  Products of $\czgate$ circuits correspond to modulo-2 vector addition in the coefficient space of \ssec{notation}; repeated gates cancel.

\subsection{Linear functions carried by an auxiliary swap walk}

Use the prefix change of variables from the proof of Theorem~6 in \cite{maslov2018shorter} on rail A, and introduce adjacent EXORs on rail B:
\begin{equation*}
\begin{aligned}
u_0&\,{=}\,0,\qquad u_j\,{=}\,a_1\,{\oplus}\,a_2\,{\oplus}\,...\,{\oplus}\,a_j,\qquad 1\,{\leq}\,j\,{\leq}\,k,\\
v_0&\,{=}\,b_1,\qquad v_j\,{=}\,b_j\,{\oplus}\,b_{j+1}\quad(1\,{\leq}\,j\,{<}\,k),\qquad v_k\,{=}\,b_k.
\end{aligned}
\end{equation*}
Thus $a_j\,{=}\,u_{j-1}\,{\oplus}\,u_j$, $b_j\,{=}\,v_0\,{\oplus}\,v_1\,{\oplus}\,...\,{\oplus}\,v_{j-1}$, and the EXOR of all $v_j$ is zero.  The functions $u_1,u_2,...,u_k$ are a basis for rail A, and any $k$ of the $k{+}1$ functions $v_0,v_1,...,v_k$ are a basis for rail B.  These functions are classical descriptions of the existing input variables.

Place $k{+}1$ labels $0,1,2,...,k$ on an auxiliary path.  At even-numbered stages, beginning at zero, swap positions $(0,1),(2,3),...$; at odd-numbered stages, swap $(1,2),(3,4),...$, omitting nonexistent pairs.  Let $\pi_t(i)$ be the label at position $i$ after $t$ stages.

Swapping positions $(j{-}1,j)$ preserves their EXOR $u_{j-1}\,{\oplus}\,u_j$ and adds it to each neighboring EXOR.  On the $v$ side, only the prefix ending at position $j{-}1$ changes, by the EXOR of its two neighboring prefixes.  These are exactly the complementary horizontal stages of \sec{ladder}.  Induction therefore gives the two parities at rung $r$ after $t$ stages:
\begin{equation}\label{eq:ladderparities}
u_{\pi_t(r-1)}\,{\oplus}\,u_{\pi_t(r)},\qquad
\bigoplus_{i=0}^{r-1}v_{\pi_t(i)},\qquad 1\,{\leq}\,r\,{\leq}\,k.
\end{equation}
By \eq{rectangle}, a rung $\czgate$ between these parities supplies a rectangle of cross-rail $\czgate$ gates.

\subsection{The rectangles exposed by the walk}

Give the initial labels cyclic coordinates modulo $k{+}1$:
\begin{equation}\label{eq:cycliclabels}
z_{2h}\,{=}\,{-}h,\qquad z_{2h+1}\,{=}\,h{+}1.
\end{equation}
The two swap matchings act on these coordinates as the reflections $z\,{\mapsto}\,1{-}z$ and $z\,{\mapsto}\,{-}z$.  Consequently,
\begin{equation}\label{eq:cyclicwalk}
z_{\pi_{2a}(i)}\,{=}\,a{+}z_i,\qquad
z_{\pi_{2a+1}(i)}\,{=}\,a{+}1{-}z_i.
\end{equation}
The first $r$ labels in \eq{cycliclabels} form a cyclic interval: $[-h,h]$ for $r\,{=}\,2h{+}1$, and $[1{-}h,h]$ for $r\,{=}\,2h$.  Rotations and reflections therefore expose every proper nonempty cyclic interval.

From now on, subscripts on $u,v$ denote these cyclic coordinates.  Put
\begin{equation*}
c_z\,{=}\,u_z\,{\oplus}\,u_{z+1}.
\end{equation*}
The EXOR of all $c_z$ is zero, and $c_0,c_1,...,c_{k-1}$ form a basis for rail A.  Indeed, they recover all $u_z$ successively, starting from $u_0\,{=}\,0$.  The variables $v_0,v_1,...,v_{k-1}$ likewise form a basis for rail B.

For a prefix interval $[i,j]$, the two $u$ labels adjacent to its boundary in \eq{ladderparities} are either $u_i,u_{j+1}$ or $u_{i-1},u_j$.  Their EXORs are, correspondingly, $c_i\,{\oplus}\,c_{i+1}\,{\oplus}\,...\,{\oplus}\,c_j$ and $c_{i-1}\,{\oplus}\,c_i\,{\oplus}\,...\,{\oplus}\,c_{j-1}$.  The rung slots over the $2(k{+}1)$ states in \eq{cyclicwalk} thus give both families of interval rectangles
\begin{equation}\label{eq:intervalrectangles}
R_\delta[i,j]\,{:=}\,
\czgate\left(\bigoplus_{z=i}^{j}c_{z+\delta},\
                 \bigoplus_{z=i}^{j}v_z\right),\qquad \delta\,{=}\,0,{-}1.
\end{equation}
Here $[i,j]$ is a proper nonempty cyclic interval, with all subscripts reduced modulo $k{+}1$.  An empty interval denotes the identity circuit.

Only the sampled physical boundaries $1,2,...,k{+}1$ are needed.  Substitution in \eq{cyclicwalk} gives
\begin{equation}\label{eq:walkreverse}
\pi_{t+k+1}(i)\,{=}\,\pi_t(k{-}i).
\end{equation}
For example, the identities used are $z_{k-i}\,{=}\,\frac{k+1}{2}{+}z_i$ for odd $k$, and $z_{k-i}\,{=}\,\frac{k}{2}{+}1{-}z_i$ for even $k$.  Reversal exchanges the two boundary labels and replaces the $v$ prefix by its complement.  The EXOR of all $v_z$ is zero, so the complementary prefix has the same value.  Hence
\begin{equation*}
M_{t+k+1,r}\,{=}\,M_{t,k+1-r}.
\end{equation*}
The rung sets at states zero and $k{+}1$ are the same, and states $1,2,...,k{+}1$ already expose every rectangle in \eq{intervalrectangles}.  At $t\,{=}\,0$, \eq{walkreverse} also proves that the auxiliary order is reversed after $k{+}1$ stages.  Substituting in \eq{ladderparities} gives $A_{k+1}\,{=}\,B_{k+1}\,{=}\,J_k$.

\subsection{Canceling rectangles to obtain individual gates}

Arrange the cells, $\czgate(c_i,v_j)$, we want to obtain along the diagonals as follows: main diagonal, first superdiagonal, first subdiagonal, second superdiagonal, second subdiagonal, etc.  Place $R_0[i,j]$ at position $(i,j)$, for $0\,{\leq}\,i\,{\leq}\,j\,{<}\,k$,
and $R_{-1}[i+1,j]$ at position $(j,i)$, for $0\,{\leq}\,i\,{<}\,j\,{<}\,k$. Arrange the $R[\cdot,\cdot]$ squares we have by diagonals: main diagonal, first subdiagonal, first superdiagonal, second subdiagonal, second superdiagonal, etc.  Notice that each list has $2k{-}1$ items, and the first $i$ items of the first list are constructible using the first $i$ items of the second list, by incrementing $i$. Each square contains exactly one cell on the corresponding new diagonal; all its other cells are already individually constructible and can be canceled. Different squares supply different cells of the new diagonal.  Thus all individual $\czgate(c_i,v_j)$ gates are obtainable.

The following two figures graphically illustrate which rectangles are generated and how they are used to obtain individual $\czgate(c_i,v_j)$.  The rules of the game here are simple: given the specified $k^2$ patches of blue and yellow colors, find their modulo-2 overlap that generates an arbitrary $1{\times} 1$ cell.

\begin{figure}[p]
\centering
\includegraphics[width=0.92\textwidth]{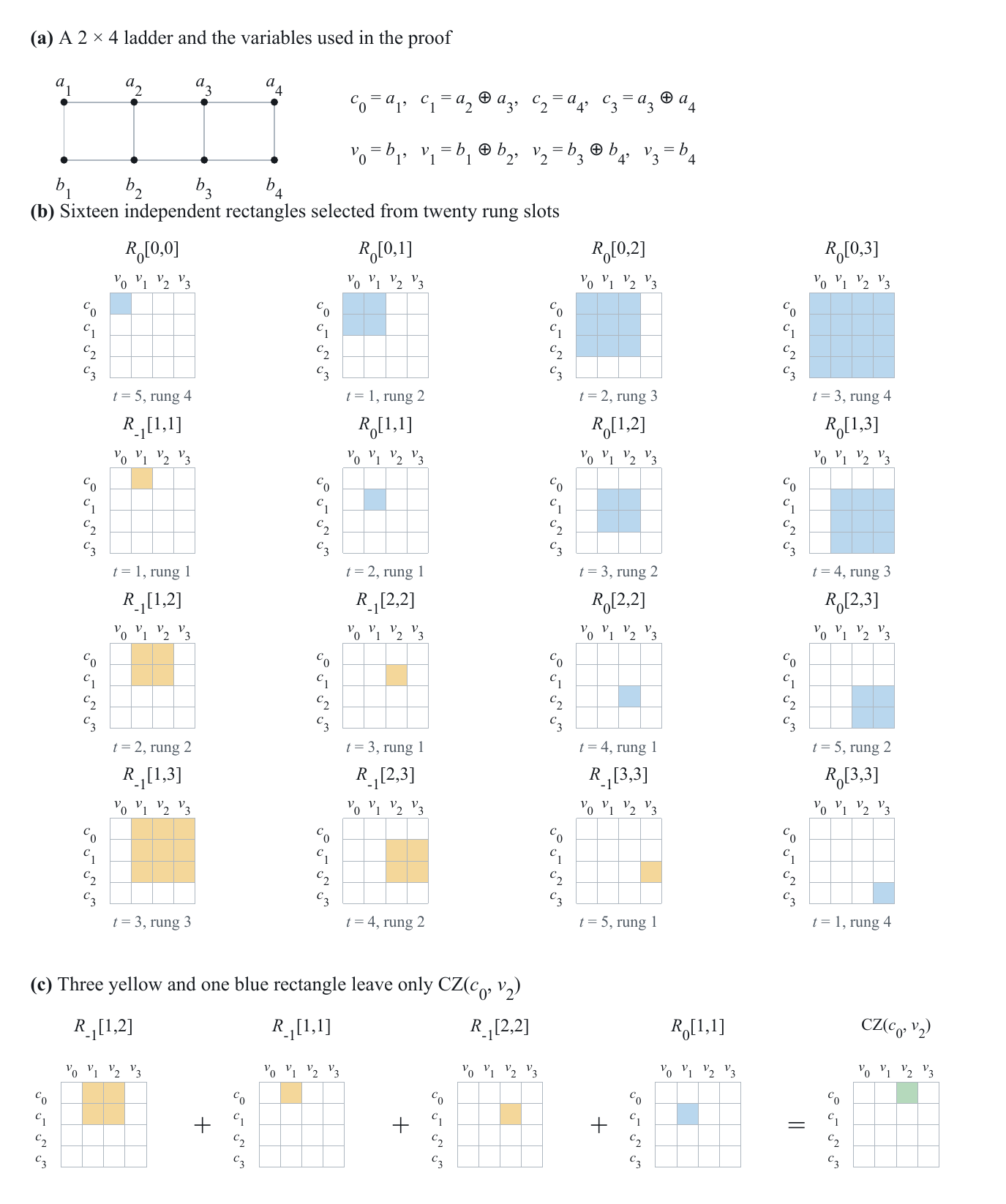}
\caption{Rectangles generated on a $2{\times}4$ ladder.  (a) The physical ladder and the change of coordinates.  (b) Sixteen independent rectangles selected from the twenty rung slots, in the regular triangular arrangement.  Each shaded cell denotes the $\czgate$ gate between its row and column variables.  Blue denotes $R_0[i,j]$ and yellow denotes $R_{-1}[i,j]$, as defined in \eq{intervalrectangles}.  The labels give the number $t$ of completed horizontal stages and the rung, numbered from left to right.  (c) The three yellow rectangles $R_{-1}[1,2]$, $R_{-1}[1,1]$, and $R_{-1}[2,2]$, together with the blue rectangle $R_0[1,1]$ generate $\czgate(c_0,v_2)$, shown as one green cell.  Overlap is taken over $\FF_2$; i.e., pairs of cells cancel.}
\label{fig:ladderrectangles}
\end{figure}

\begin{figure}[p]
\centering
\includegraphics[width=0.92\textwidth]{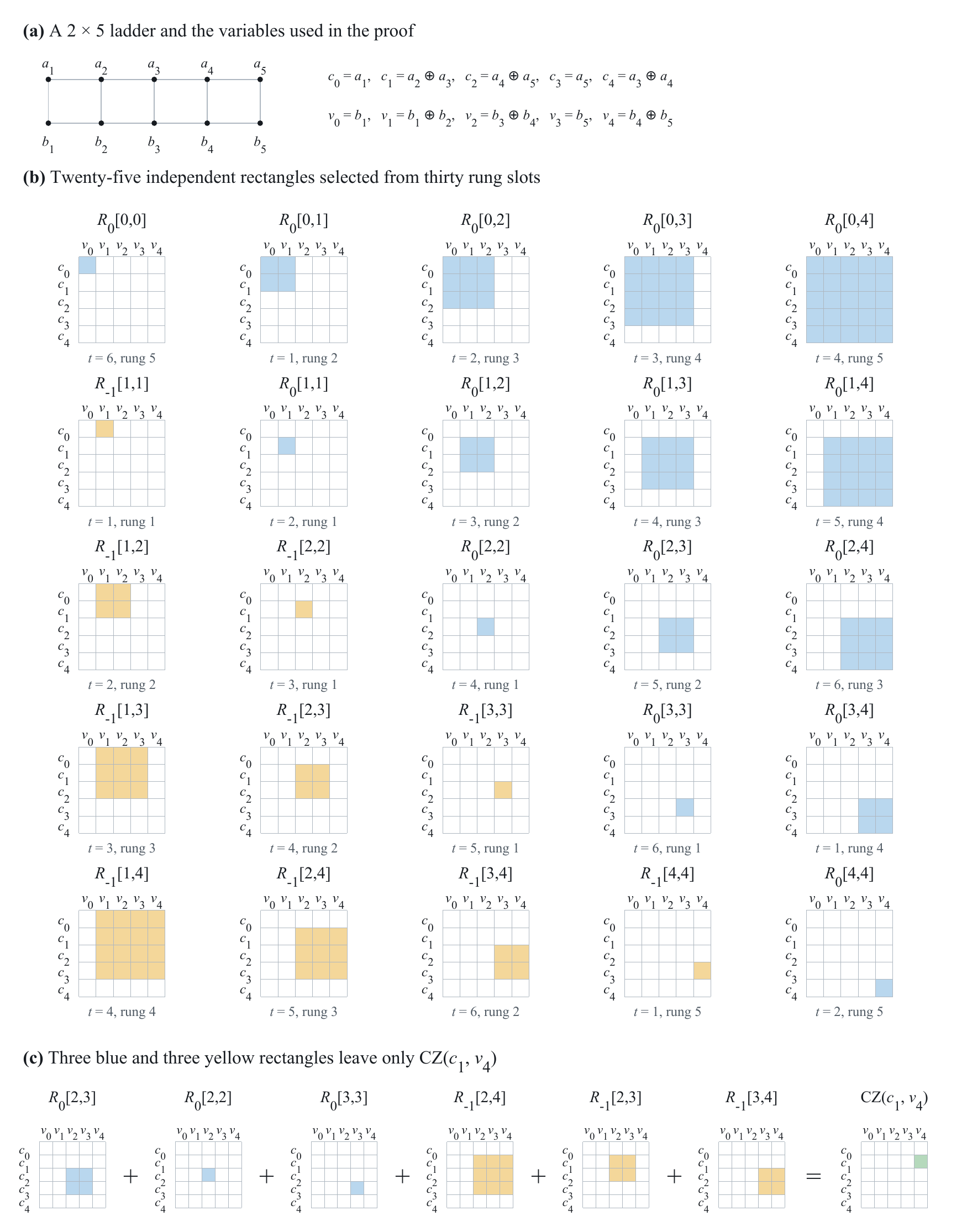}
\caption{A $2{\times}5$ example, using the same colors and notation as \fig{ladderrectangles}.  (a) The ladder and the changed variables.  (b) Twenty-five independent rectangles selected from the thirty rung slots.  (c) A configuration of six rectangles leaves only $\czgate(c_1,v_4)$: the three blue rectangles $R_0[2,3]$, $R_0[2,2]$, and $R_0[3,3]$, and the three yellow rectangles $R_{-1}[2,4]$, $R_{-1}[2,3]$, and $R_{-1}[3,4]$.}
\label{fig:ladderrectanglesfive}
\end{figure}

\clearpage

\section{Explicit small circuits and phase programs}\label{app:small}

We give two short positive witnesses for \ssec{small}.  Qubits are numbered by rows, beginning with one.  An ordered pair $(c,t)$ denotes $\cnotgate(c,t)$, and braces contain gates in one matching layer.  The $\cnotgate$ layers are executed in the listed order.  Input powers of $\pgate$ are available for the linear correction described below.

\subsection{Four qubits}

Place qubits $1,2$ on the top row and $3,4$ on the bottom row of a $2\,{\times}\,2$ square.  Use
\begin{equation}
\begin{split}
L_1&\,{=}\,\{(1,3),(2,4)\},\qquad L_2\,{=}\,\{(1,2),(3,4)\},\\
L_3&\,{=}\,\{(3,1),(2,4)\},\qquad L_4\,{=}\,\{(1,2),(3,4)\}.
\end{split}\nonumber
\end{equation}
Allow Phase gates on all wires after each of the four layers.  Exact replay gives $\czgate$ rank six.  It deliberately allows nonproductive gates, which are excluded only at the saturated counting bound.

\subsection{Eight qubits}

Place qubits $1,2,3,4$ on the top row and $5,6,7,8$ on the bottom row of a $2\,{\times}\,4$ rectangle.  Put
\begin{equation}
\begin{split}
A&\,{=}\,\{(1,5),(3,2),(8,4),(6,7)\},\\
B&\,{=}\,\{(2,1),(4,3),(5,6),(7,8)\}.
\end{split}\nonumber
\end{equation}
The eight-layer circuit $(A,B,A,B,A,B,A,B)$ has $\czgate$ rank $28$.  Phase rounds after layers $2,4,6,8$ suffice, in addition to the input corrections.  This example is strictly checkerboard alternating.  It uses $32$ $\cnotgate$ gates, as compared to the universal gate-count lower bound $28$; the extra layer is forced by geometry, as established by the exhaustive exclusion.

\subsection{Extracting the program}

Let $w_1,w_2,...,w_m$ be the parities at the declared slots, and let $A$ now denote the endpoint matrix of the entire $\cnotgate$ circuit.  To prepare the desired graph state, form the phase polynomial
\begin{equation}
p_*(x)\,{=}\,2\sum_{i\,{<}\,j}\Gamma_{ij}(Ax)_i(Ax)_j\pmod{4},\nonumber
\end{equation}
where the output bits $(Ax)_i$ are evaluated over $\FF_2$.  Reduce it to its unique linear-plus-quadratic representation
\begin{equation}
p_*(x)\,{=}\,\sum_{i\,{=}\,1}^n\lambda_i x_i{+}2\sum_{i\,{<}\,j}\gamma_{ij}x_ix_j\pmod{4}.\nonumber
\end{equation}
Solve the binary system
\begin{equation}\label{eq:programsolve}
\bigoplus_{s\,{=}\,1}^m a_s q(w_s)\,{=}\,\gamma,\qquad a_s\,{\in}\,\FF_2.
\end{equation}
Apply $\pgate^{a_s}$ at slot $s$ and, at the input boundary, apply $\pgate^{b_i}$ on wire $i$, with
\begin{equation}
b_i\,{=}\,\lambda_i{-}\sum_{s\,{=}\,1}^m a_s(w_s)_i\pmod{4}.\nonumber
\end{equation}
Together, \eq{parityexpand} and \eq{programsolve} show that this program has polynomial $p_*$.  The input boundary is after the initial Hadamard stage.  The $\cnotgate$ endpoint is then absorbed exactly as in \eq{endpointabsorb}, and the final local Clifford stage supplies $L$.

\section{Geometric proof of the cycle construction}\label{app:cycles}

We prove the constructive part of \lem{cycles} by drawing and canceling $\czgate$ gates.  A chord joins the two input variables in a $\czgate$ basis vector.  Products below denote composition of $\czgate$ circuits, or modulo-2 addition of their coefficient vectors, with primary-input Phase gates omitted as in \ssec{notation}.  Thus repeated chords cancel, $\czgate(u,u)$ has zero $\czgate$ vector, and the rectangle identity \eq{rectangle} applies.  A substitution acts on both arguments of $\czgate(u,v)$.

Let $A$ be the binary matrix of the first layer of \eq{cyclelayers}, and let $R$ rotate wire values by one vertex, $(Rx)_i\,{=}\,x_{i-1}$, with indices modulo $n$.  Put $T\,{=}\,R^{-1}A$.  Describing the wire labels in a frame shifted back by one position after each layer makes every step the same substitution $T$.  The binary endpoint after $t$ layers is
\begin{equation}\label{eq:cycleprefix}
F_t\,{=}\,R^tT^t,
\end{equation}
because chronological layer $t$, beginning at zero, is $R^tAR^{-t}$.

\subsection{Even cycles}

\begin{figure}[h]
\centering
\includegraphics[width=\textwidth]{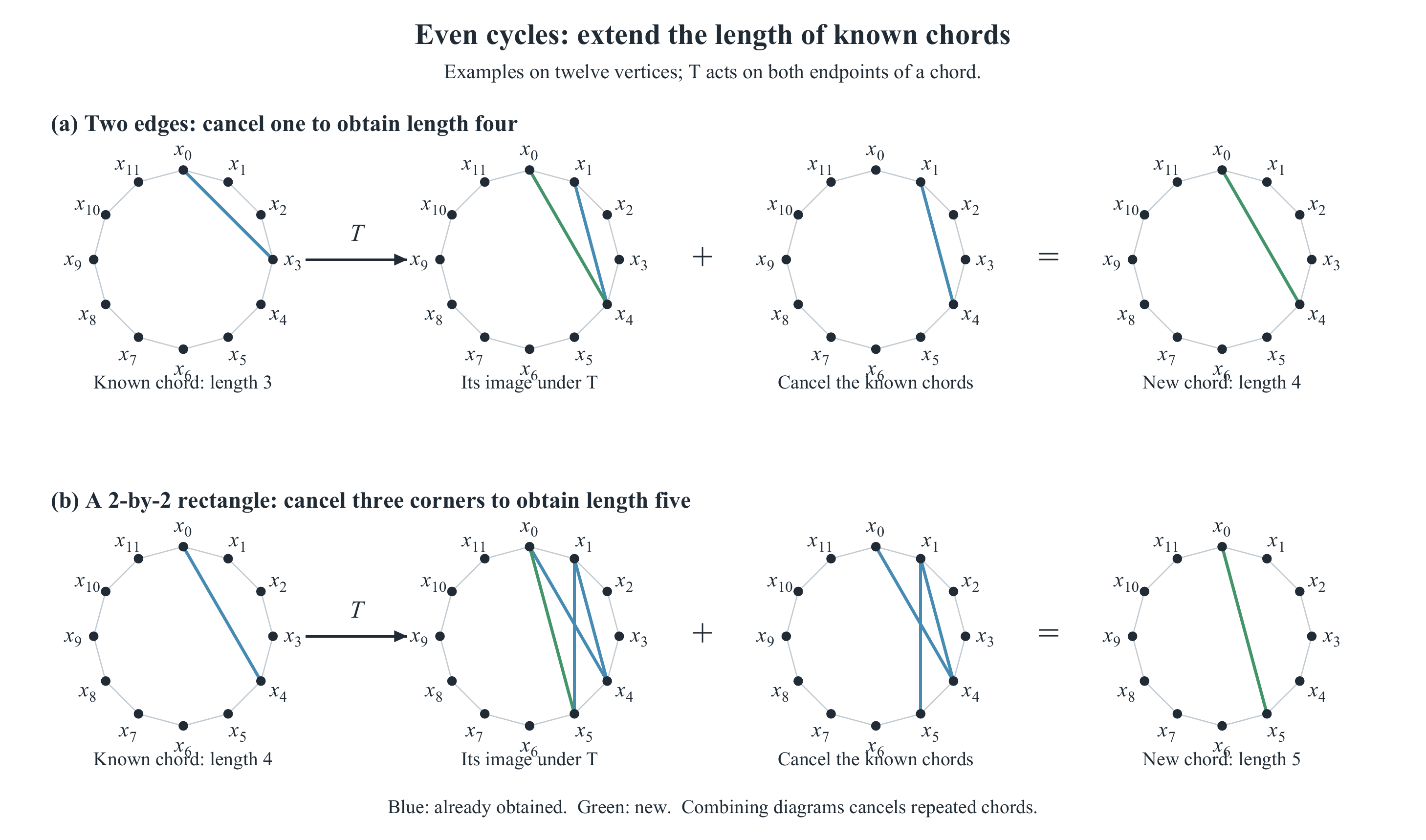}
\caption{Growing $\czgate$ chord lengths on an even cycle, illustrated for $n\,{=}\,12$.  Blue chords are already constructible; green chords are new.  (a) Applying $T$ to a length-three chord gives two edges.  Canceling the known edge leaves a length-four chord.  (b) Applying $T$ to a length-four chord gives a $2{\times}2$ rectangle.  Canceling its three known corners leaves a length-five chord.  Combining drawings means composition of $\czgate$ circuits, so repeated chords cancel.}
\label{fig:cycle-chord-growth}
\end{figure}

Let $n\,{=}\,2m$.  Draw the input variables $x_0,x_1,x_2,...,x_{n-1}$ around the cycle, and call the smaller cyclic separation of two vertices the length of their chord.  The substitution and its inverse are
\begin{equation}\label{eq:evenupdate}
\begin{aligned}
T(x_{2i})&\,{=}\,x_{2i}\,{\oplus}\,x_{2i+1},&
T(x_{2i+1})&\,{=}\,x_{2i+2},\\
T^{-1}(x_{2i})&\,{=}\,x_{2i-1},&
T^{-1}(x_{2i+1})&\,{=}\,x_{2i}\,{\oplus}\,x_{2i-1}.
\end{aligned}
\end{equation}
All subscripts are modulo $n$.  The same letter $T$ denotes its action on $\czgate$ vectors.  By \eq{increment} and \eq{cycleprefix}, the increments in chronological layer $t$ are
\begin{equation}\label{eq:evenincrements}
T^t\czgate(x_{2i},x_{2i+1}),\qquad i\,{=}\,0,1,2,...,m{-}1.
\end{equation}
For later use in \lem{halfturn}, \eq{evenupdate} also gives the binary matrix identity
\begin{equation}\label{eq:cyclecellshift}
T^2\,{\oplus}\,T\,{=}\,R^{-2}.
\end{equation}
Indeed, applying $T$ twice and adding its single application shifts every input variable forward by two positions.

Say that a chord is available in $[l,h]$ if its vector is a binary sum of increments in \eq{evenincrements} with $l\,{\leq}\,t\,{\leq}\,h$.  We temporarily allow negative times; a common invertible substitution will translate the entire interval to physical layers.  Applying $T$ shifts an interval forward by one, and applying $T^{-1}$ shifts it back by one.  Every even starting vertex is present in every layer, so shifting all subscripts by two does not change availability.

\begin{figure}[t]
\centering
\includegraphics[width=\textwidth]{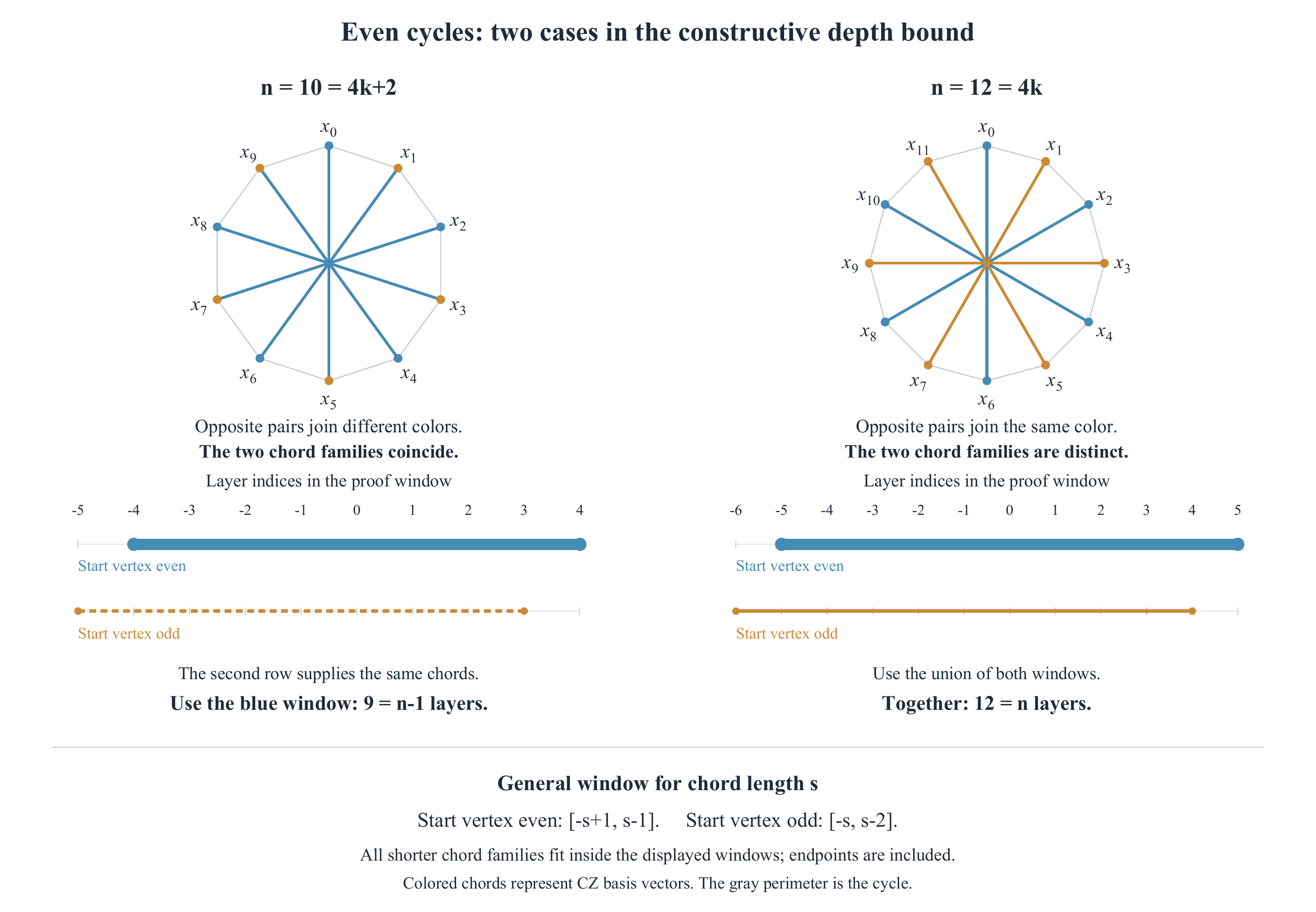}
\caption{The final chord families and their layer intervals in the even-cycle construction.  Vertices are colored by parity.  For $n\,{=}\,10\,{=}\,4k{+}2$, every opposite pair has one endpoint of each color, so the two starting-parity families coincide.  The blue interval alone has $n{-}1$ layers and contains every shorter-chord interval.  For $n\,{=}\,12\,{=}\,4k$, the blue and orange opposite pairs are distinct families; the union of their intervals has $n$ layers.  Interval endpoints are included.  The negative indices describe availability in the proof and are shifted together to nonnegative physical layer indices.}
\label{fig:cycle-chord-windows}
\end{figure}

We construct chords in increasing order of length $s$.  There are two families, according to whether the starting vertex is even or odd, with availability intervals
\begin{equation}\label{eq:cycleintervals}
\begin{array}{c|c}
\text{chord},\quad 1\,{\leq}\,s\,{\leq}\,m&\text{available times}\\\hline
\czgate(x_{2i},x_{2i+s})&[{-}s{+}1,\,s{-}1]\\
\czgate(x_{2i+1},x_{2i+1+s})&[{-}s,\,s{-}2].
\end{array}
\end{equation}
For $s\,{=}\,1$, these are the starting chords at time zero and their images under $T^{-1}$ at time $-1$.

To obtain the next even length $s$, apply $T$ to an even-start chord of length $s{-}1$.  By \eq{evenupdate}, this gives two edges:
\begin{equation*}
T\czgate(x_{2i},x_{2i+s-1})
\,{=}\,\czgate(x_{2i},x_{2i+s})\,\czgate(x_{2i+1},x_{2i+s}).
\end{equation*}
The second edge has length $s{-}1$ and is already constructible.  Cancel it to obtain the first, of length $s$.

For the next odd length $s$, the same substitution gives a $2{\times}2$ rectangle:
\begin{equation*}
\begin{aligned}
T\czgate(x_{2i},x_{2i+s-1})
&\,{=}\,\czgate(x_{2i},x_{2i+s-1})\,\czgate(x_{2i},x_{2i+s})\\
&\qquad\czgate(x_{2i+1},x_{2i+s-1})\,\czgate(x_{2i+1},x_{2i+s}).
\end{aligned}
\end{equation*}
Its corners have lengths $s{-}1,s,s{-}2,s{-}1$.  Canceling the three shorter chords leaves the desired length-$s$ chord.  \fig{cycle-chord-growth} illustrates both cancellations.

This constructs the even-start family.  Applying $T^{-1}$ to its chord starting at $2i{+}2$ gives the odd-start chord of the same length.  When $s$ is odd, it also gives one chord of length $s{-}1$, which is already constructible and is canceled.  The even-start construction combines the preceding even-start interval shifted forward by one with shorter-chord intervals.  Their union is $[{-}s{+}1,s{-}1]$.  Shifting back by one gives $[{-}s,s{-}2]$ for the odd-start family; its possible shorter correction fits inside that interval.  This proves \eq{cycleintervals} by induction.

At the final length $s\,{=}\,m$, the chords join opposite vertices.  If $m$ is odd, these vertices have different parity, so the two families in \eq{cycleintervals} are the same set of chords, described from opposite ends.  We retain only the even-start interval $[{-}m{+}1,m{-}1]$; all shorter-chord intervals fit inside it.  Applying $T^{m-1}$ translates this interval to $[0,2m{-}2]$.  Thus the first $n{-}1$ physical layers span all $D(n)$ $\czgate$ dimensions for $n\,{=}\,4k{+}2$.  They contain exactly $D(n)$ gates, so their increments form a basis.

If $m$ is even, opposite vertices have the same parity.  The even-start and odd-start families of opposite chords are distinct, and we use both intervals.  Their union, $[{-}m,m{-}1]$, contains all shorter-chord intervals.  Applying $T^m$ translates it to $[0,2m{-}1]$, giving depth $n$ for $n\,{=}\,4k$.  \fig{cycle-chord-windows} illustrates the distinction.  The separate matching argument in \ssec{cycles} proves that the additional layer for $n\,{=}\,4k$ is necessary.

\subsection{Odd cycles}

Let $n\,{=}\,2m{+}1$.  The rotating description gives
\begin{equation*}
T(x_i)\,{=}\,\begin{cases}
x_{i+1},&i\text{ even},\\
x_i\,{\oplus}\,x_{i+1},&i\text{ odd},
\end{cases}
\end{equation*}
where subscripts are modulo $n$.  On linear functions put $S\,{=}\,T^2\,{\oplus}\,T$.  Direct substitution gives
\begin{equation}\label{eq:oddshift}
S(x_i)\,{=}\,x_{i+2}\quad(0\,{\leq}\,i\,{<}\,n{-}1),\qquad
S(x_{n-1})\,{=}\,x_0\,{\oplus}\,x_1.
\end{equation}
As for $T$, the action of $S$ on a $\czgate$ vector substitutes in both arguments.

\begin{figure}[t]
\centering
\includegraphics[width=\textwidth]{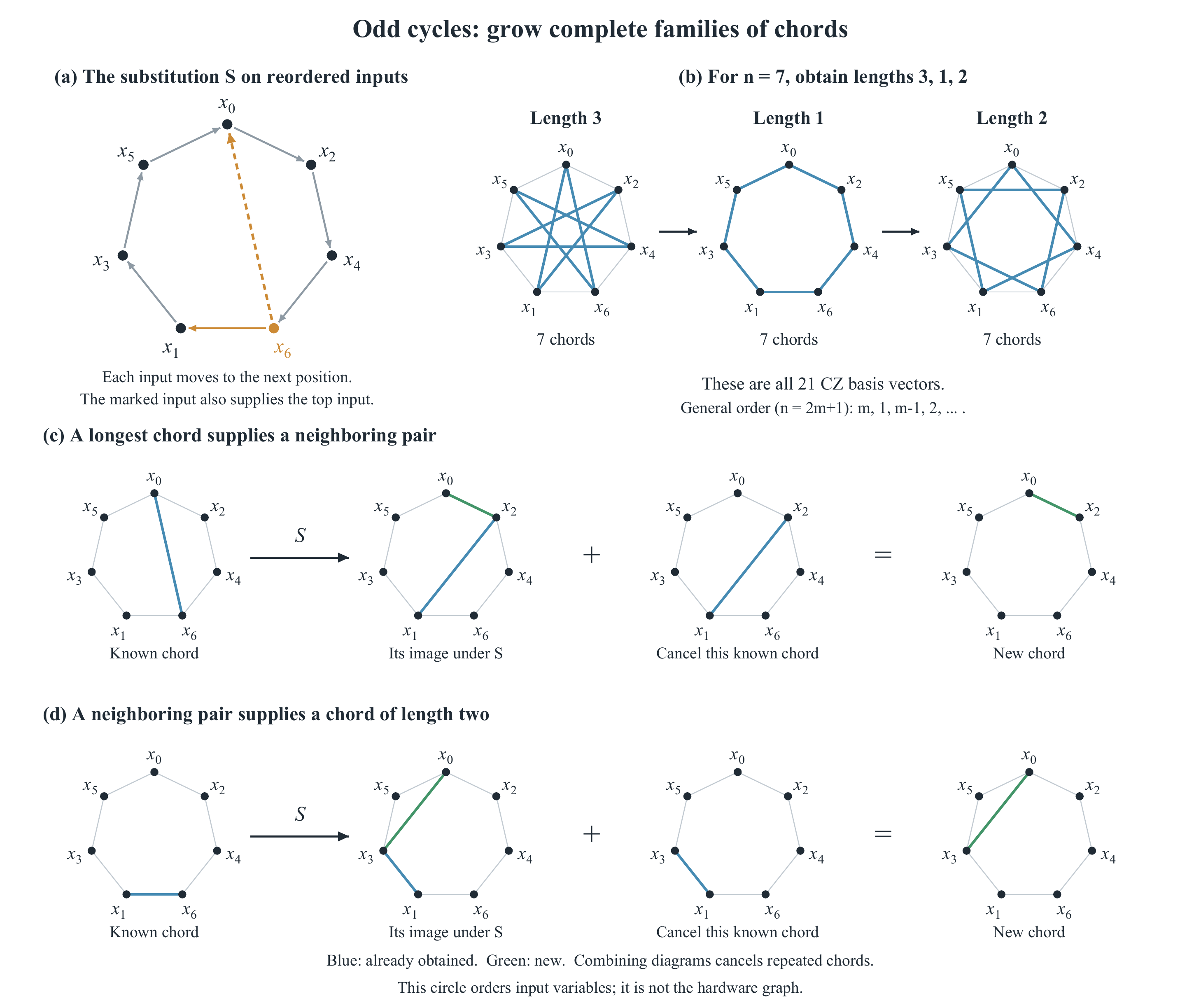}
\caption{Geometric proof for odd cycles, illustrated for $n\,{=}\,7$.  The circles show the reordered input variables used in the proof.  (a) The substitution $S$ advances each variable one position, with one additional arrow from $x_6$ to $x_0$.  (b) Constructing the three chord-length families $3,1,2$ supplies all $21$ $\czgate$ basis vectors.  In general, the order is $m,1,m{-}1,2,m{-}2,3,...$ for $n\,{=}\,2m{+}1$.  (c) Applying $S$ to a known longest chord and canceling another known longest chord leaves a neighboring pair.  (d) Applying the same cancellation to a known neighboring pair leaves a length-two chord.  Blue chords are already constructible and green chords are new; repeated chords cancel.}
\label{fig:odd-cycle-chords}
\end{figure}

Draw the original input variables in the cyclic order
\begin{equation*}
x_0,x_2,x_4,...,x_{2m},x_1,x_3,...,x_{2m-1}.
\end{equation*}
This circle records the order of variables used in the proof.  By \eq{oddshift}, $S$ moves every variable one position forward, except that $x_{2m}$ supplies both the next variable $x_1$ and the first variable $x_0$; see \fig{odd-cycle-chords}(a).  The inverse moves every variable one position back, except that $S^{-1}(x_1)\,{=}\,x_{2m}\,{\oplus}\,x_{2m-1}$.  Tracing $m$ inverse steps gives $x_{i+1}$ for even $i$ and $x_i\,{\oplus}\,x_{i+1}$ for odd $i$, exactly the update $T$.  Therefore
\begin{equation}\label{eq:oddcycleidentity}
S^mT\,{=}\,I,\qquad T\,{=}\,S^{-m}.
\end{equation}

Consider the sequence of $\czgate$ vectors
\begin{equation*}
E_k\,{=}\,\czgate(S^k x_0,S^k x_1),\qquad k\,{\in}\,\ZZ.
\end{equation*}
Its span is preserved by $S$ and $S^{-1}$.  We show that it contains every chord of the reordered circle.  All lengths below are measured in that circle.

First, $E_0$ is a longest chord, of length $m$.  Its first $m$ forward and $m$ backward images give all $n$ chords of length $m$ individually.  There is no extra chord in the forward steps.  The first backward step has a repeated-variable term, which has zero $\czgate$ vector, and the remaining backward steps are ordinary rotations.

Next, a single chord of a new length can be spread around the circle.  To see this, number the positions $0,1,2,...,2m$ in the displayed order and describe a chord by its endpoint positions.  Let $1\,{\leq}\,r\,{\leq}\,m$.  For a chord $(0,r)$, rotate $m$ steps forward and $m$ steps backward, canceling any known correction at each step.  The only extra chord occurs in the forward steps and is $(0,m{+}1{-}r)$, of length $m{+}1{-}r$.  Thus $(0,r)$ supplies its entire length-$r$ family if that correction family is already known.  For a chord $(0,n{-}r)$, the only extra chord occurs in the backward steps and is $(m{+}r,2m)$, of length $m{-}r$.  Its entire family is therefore obtained if length $m{-}r$ is already known.  For $r\,{=}\,m$, this last correction has identical endpoints and vanishes.  In either case, the forward and backward walks together cover all $n$ rotations of the chord.

Suppose now that all chords of length $\ell$ are known.  Apply $S$ to the chord $(m,m{-}\ell)$.  Its image is its ordinary rotation together with $(0,m{+}1{-}\ell)$.  Cancel the known rotation to leave a chord of length $m{+}1{-}\ell$.  The preceding rotation argument supplies every chord of this new length, using only length-$\ell$ corrections.  Alternatively, applying $S$ to $(m,m{+}\ell)$ and canceling its known rotation leaves $(0,m{+}1{+}\ell)$, of length $m{-}\ell$; again length-$\ell$ corrections suffice.  Starting with the longest chords and alternating these two cancellations gives the lengths
\begin{equation*}
m,1,m{-}1,2,m{-}2,3,...
\end{equation*}
until every length $1,2,3,...,m$ has been constructed.  Each length has $n$ different chords.  These are all $mn\,{=}\,D(n)$ $\czgate$ basis vectors.

For example, when $n\,{=}\,7$, the order of lengths is $3,1,2$.  The two steps are
\begin{equation*}
\begin{aligned}
S\czgate(x_0,x_6)&\,{=}\,\czgate(x_2,x_1)\,\czgate(x_0,x_2),\\
S\czgate(x_6,x_1)&\,{=}\,\czgate(x_1,x_3)\,\czgate(x_0,x_3).
\end{aligned}
\end{equation*}
In the first line, cancel the known length-three chord to obtain a length-one chord.  After constructing that entire family, the second line gives a length-two chord.  \fig{odd-cycle-chords} illustrates the construction of all $21$ $\czgate$ basis vectors.

It remains to count physical layers.  By \eq{oddshift}, $S^jE_0\,{=}\,\czgate(x_{2j},x_{2j+1})$ for $j\,{=}\,0,1,2,...,m{-}1$.  Applying $T$ gives the first-layer increments, which join $(1,2),(3,4),...,(n{-}2,n{-}1)$.  Each further layer applies a common $T$.  Using \eq{oddcycleidentity}, the increments in layer $t$ are
\begin{equation}\label{eq:oddincrements}
E_{j-m(t+1)},\qquad j\,{=}\,0,1,2,...,m{-}1.
\end{equation}
The first $n$ layers in \eq{oddincrements} give exactly $mn\,{=}\,D(n)$ consecutive vectors,
\begin{equation*}
E_{-mn},E_{-mn+1},...,E_{-1}.
\end{equation*}
They are independent.  Otherwise, the first dependence in this block would make its preceding independent segment span a subspace preserved by $S$.  Since $S$ is invertible, this subspace is also preserved by $S^{-1}$ and would contain the entire sequence, contradicting its full span just proved.  This establishes depth $n$ for every odd cycle.

\end{document}